\documentclass[twocolumn,english,aps,superscriptaddress,prb,floatfix,longbibliography]{revtex4-2}
\usepackage[latin9]{inputenc}
\usepackage{color}
\usepackage{babel}
\usepackage{amsmath}
\usepackage{amssymb}
\usepackage{graphicx}
\usepackage{esint}
\usepackage[unicode=true,pdfusetitle,
 bookmarks=true,bookmarksnumbered=false,bookmarksopen=false,
 breaklinks=true,pdfborder={0 0 1},backref=false,colorlinks=true]
 {hyperref}
\hypersetup{
 pdfborderstyle=,linkcolor=blue,citecolor=blue,urlcolor=blue}

\makeatletter

\providecommand{\tabularnewline}{\\}

\usepackage{babel}

\usepackage{color}

\usepackage{color}
\usepackage{dsfont}
\usepackage{braket}

\makeatother

\begin{document}
\title{Fragile magnetic order in metallic quasicrystals}
\author{Ronaldo N. Araújo}
\affiliation{Instituto de Física de São Carlos, Universidade de São Paulo, 13560-970,
São Carlos, SP Brazil}
\author{Carlo C. Bellinati}
\affiliation{Instituto de Física, Universidade de São Paulo, 05315-970, São Paulo,
SP, Brazil}
\author{Eric C. Andrade}
\affiliation{Instituto de Física, Universidade de São Paulo, 05315-970, São Paulo,
SP, Brazil}
\begin{abstract}
Inspired by recent experimental studies of local magnetic moments
interacting with a metallic quasicrystal, we study the low-temperature
fate of spins placed in two-dimensional tilings. In the diluted
local moment limit, we calculate the spin relaxation rate $1/T_{1}$,
as measured by electron spin resonance, and show that it displays
a marked dependence on the system size $N$ and the filing $n$ of the electronic bath.  For a finite concentration of spins, we integrate
out the conduction electrons and generate an effective magnetic coupling
between the local moments, which we treat as Ising spins.  Despite
the strongly frustrating nature of the magnetic couplings and the
lack of periodicity in the problem, we find long-range orders for
finite $N$ in our large-scale Monte Carlo simulations. However, the
resulting magnetically state is fragile, as clusters of essentially
free spins fluctuate down to very low temperatures. 
\end{abstract}
\date{\today}

\maketitle

\section{Introduction}

Quasicrystals are ordered solids with a peculiar non-periodic arrangement
of atoms: their diffraction patterns show well-defined Bragg peaks
while exhibiting symmetries that are forbidden in periodic systems
\citep{shechtman84,janssen18}. Over the last years, platforms based
on Moiré systems \citep{ahn18,yao18,uri23}, optical lattices \citep{viebahn19},
and arrays of artificial atoms \citep{kempkes19} are complementing
the existing solid-state materials and opening the venues for new
experimental and theoretical investigations on quasicrystals. The
lack of periodicity in quasicrystals invalidates Bloch's theorem and
poses an extra challenge in understanding the effects of topology
and electronic correlations in a quasiperiodic environment with both
conventional and unconventional behavior reported, for instance, for
topological electronic properties \citep{huang18,ghadimi21, fonseca23}, local
moment magnetism \citep{deguchi12,takemura15,andrade15,hartman16,koga17}
and superconductivity \citep{sakai17,kamiya18,araujo19,sakai19,rai20,cao20}.

One interesting question in metallic quasicrystals is the resulting ordered
state of local moments placed on this system as we integrate out the conduction
electrons. For its crystalline counterpart, these local moments interact
via the Ruderman--Kittel--Kasuya--Yosida (RKKY) coupling \citep{fazekas99}.
Due to the underlying lattice periodicity, the local moment's position
is random, and for a moderate concentration of magnetic impurities,
one typically finds a spin-glass ground state \citep{fischer93}.
In quasicrystals, the situation is distinct. First, the RKKY interactions
are more complex because of the novel electronic structure in quasiperiodic
systems \citep{kohmoto86,arai88,fujiwara89,fujiwara91,jazbec14, jagannathan21}.
Second, one may envision a non-random placement of the local moments:
quasicrystals are characterized by a set of local environments arranged
in a non-periodic fashion, which can be labeled by the number of nearest neighbors
$z$. It is then reasonable to assume that it would be energetically
more favorable for the local moments to sit at a particular local
environment, which could favor ordered magnetic states over the glassy
state \citep{thiem15a,thiem15b,miyazaki20}.

The majority of experiments identify spin-glass order in quasicrystals
\citep{fisher99,goldman13,kong14,cabrera-baez19,shiino21}. However,
important differences are present in the periodic approximations to
the quasicrystals, known as approximants, where long-range magnetic
order is often reported \citep{goldman13,cabrera-baez19,shiino21},
indicating a non-trivial role of the system unit-cell size.

We follow Refs. \citep{thiem15a,thiem15b} to study the ordering of
local moments placed in a magnetic quasicrystal and model the local
moments as Ising spins for convenience.  We work mainly on the square
approximants of the two-dimensional octagonal or Ammann-Beenker tiling
\citep{socolar89,anu06,deneau89},  but show that our results also hold for the Penrose mofit.
We assume the spins occupy a particular local environment of the quasicrystal,  determined by the number of nearest neighbors
$z$,  and interact via an average RKKY interaction \citep{duneau91},
which we compute using the Kernel Polynomial Method (KPM) \citep{weisse06}.
Using large-scale Monte-Carlo (MC) simulations, we find long-ranged
magnetic order for most of the parameters considered. However, the
order is fragile: a fraction of the sites experience a zero exchange
field and keep fluctuating down to $T\to0$, implying an extensive
ground state entropy and enhanced sensitivity of this magnetic state
towards perturbations.

In the complementary limit of diluted spins, we investigate spin-flip
relaxation rate between the local magnetic moments and the conduction
electrons at the Fermi level. This quantity, measured via electron
spin resonance (ESR) \citep{barnes81}, gives direct access to the
local quasicrystalline electronic environment. We compute its evolution
with the approximant size and show that it may increase or decrease,
depending on the position of the electrons' Fermi level. Applying
our results to recent experimental data \citep{cabrera-baez19}, we
conclude the Fermi energy of the conduction electrons is most likely
located close to a pseudogap in the density of states of the quasicrystal,  because the relaxation rate diminishes as we go from the approximant to the quasicrystal. 

The paper is organized as follows. In Sec. \ref{sec:electronic},
we discuss the electronic properties of the Ammann-Beenker tiling
and the resulting RKKY interactions calculated via the KPM. In Sec.
\ref{sec:order}, we discuss the magnetic model for Ising spins placed
on a given local environment of this tiling. We perform large-scale
Monte simulations to characterize the resulting magnetically ordered
states and discuss their peculiarities. In Sec. \ref{sec:penrose}
we briefly discuss that our scenario also holds in the Penrose lattice.
In Sec. \ref{sec:epr}, we discuss the spin-flip relaxation rate when
the local moments are coupled to the quasicrystalline electronic bath
in the limit of diluted impurities. Finally, we connect our results
to experimental results and conclude the paper in Sec. \ref{sec:conclusions}.
In Appendix \ref{sec:further}, we discuss the robustness of our minimal
magnetic model with respect to extra exchange terms. 


\section{\label{sec:electronic}Electronic states and magnetic couplings}

\subsection{Tiling}

\begin{figure}[t]
\centering{}\includegraphics[width=1\columnwidth]{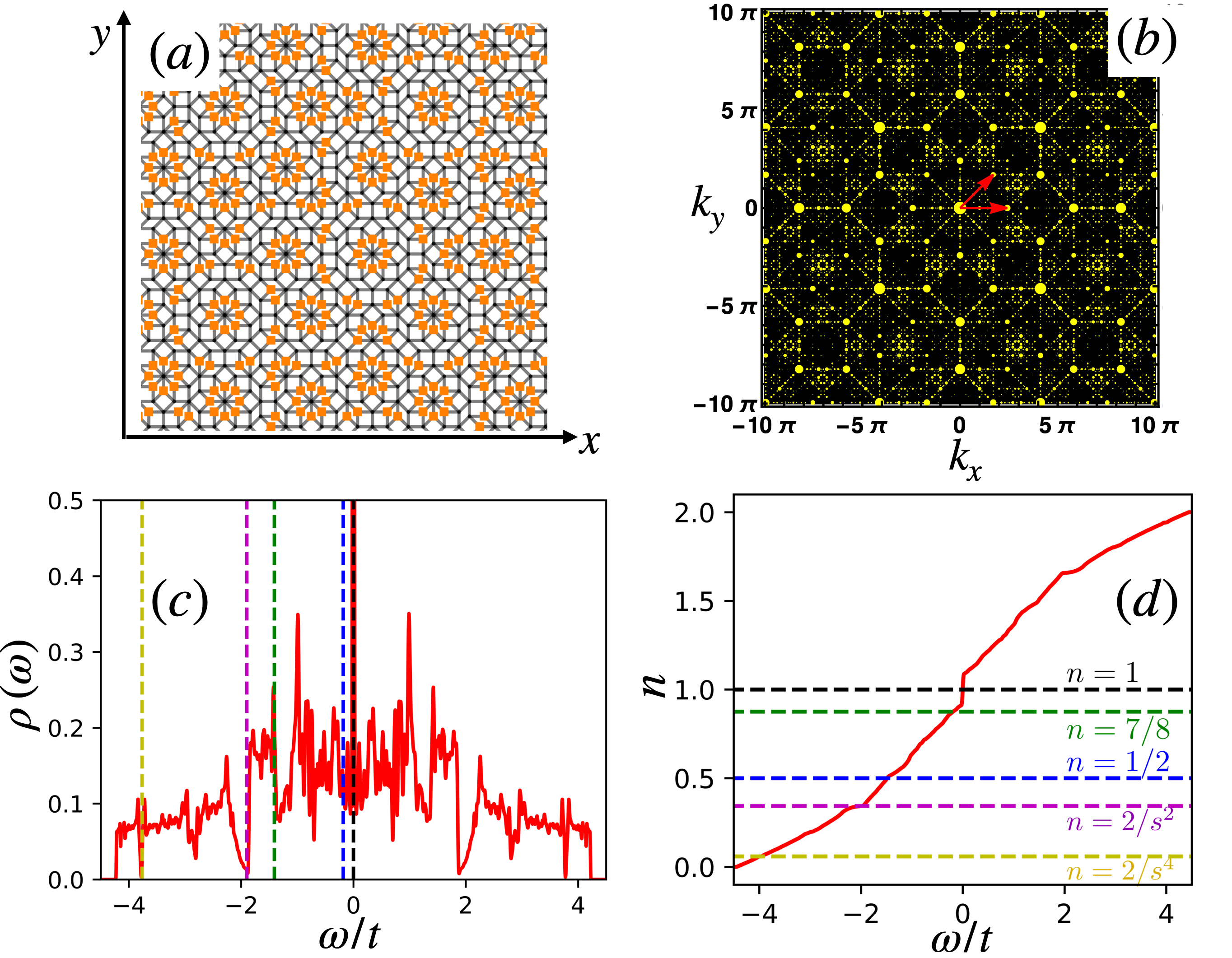}\caption{\label{fig:tiling_def}Structural and electronic properties of the
two-dimensional tiling. (a) The Ammann-Beenker tiling. The squares
highlight sites with coordination number $z=3$; (b) Structure factor
for the $N=8119$ approximant considering only the sites with $z=3$.
The width of the dots is proportional to their intensity. The arrows
correspond to the reciprocal vectors $\pi\left(1,0\right)$ and $\pi\left(1/\sqrt{2},1/\sqrt{2}\right)$,
which originate at the $\Gamma$ point; (c) Density of states, $\rho\left(\omega\right)$,
as a function of the energy, $\omega/t$, for the $N=1607521$ approximant
considering nearest-neighbor hoping $t$ only; (d) Electronic filing as
a function of $\omega/t$. The horizontal dashed lines highlight the
filings $n$ considered in this work. $s=1+\sqrt{2}$ is the silver ratio; }
\end{figure}

As a minimal model for a quasicrystal, we consider the Ammann-Beenker
tiling shown in Fig. \ref{fig:tiling_def}(a). This tiling contains
$45^{\circ}$ rhombuses and squares as building blocks and six distinct
local environments with coordination numbers $z=3,\cdot\cdot\cdot,8.$
In Fig. \ref{fig:tiling_def}(a), we highlight the sites with $z=3$.
These environments correspond to the following fraction of the total
number of sites $N$ in the system $f_{z}$: $f_{3}=1/s$, $f_{4}=2/s^{2}$,
$f_{5}=2/s^{3}$, $f_{6}=2/s^{4}$, $f_{7}=1/s^{5}$, and $f_{8}=1/s^{4}$,
where $s=1+\sqrt{2}$ is the silver ratio. We follow Ref. \citep{deneau89}
and generate square approximants to this tiling with $N=239,1393,8119,47321,275807,1607521$
sites,  and linear size $L_k=s^k$, with the integer $k$ ranging from $3$ in the approximant $N=239$  to $8$ in the mosaic $N=1607521$.

Experimentally,  an approximant is the periodic repetition of a unit cell containing a finite portion of the quasicrystal.  The unit cell size can be taken as the effective size of the approximant since all quasiperiodicity comes from it.  For an infinite-sized unit cell,  one recovers the true quasicrystal.  In this work, we will consider finite tilings with open boundary conditions as our approximants to the true quasicrystal. In principle, different constructions are possible, for instance, taking small subsets of the quasicrystal and repeating them periodically,  a situation closer to the experimental one.  Periodic boundary conditions usually lead to a faster approach to the thermodynamic limit because they avoid a variable coordination number at the sample boundaries, for instance.  In quasicrystals, we have an intrinsic distribution of $z$ and employing periodic boundary conditions does not necessarily lead to a faster convergence to the thermodynamic limit.  Moreover, as we will show in Sec. ~\ref{sec:order}, we recover the usual finite size scaling of thermodynamic quantities using our definition of approximants with open boundary conditions ~\citep{sorensen91,ledue95}.

We focus on the most frequent local environments of the Ammann-Beenker tiling,  those with  $z=3$ and $z=4$.  Moreover,  they can be related to other $z$ in different approximants by the inflation/deflation symmetry: selecting a subset of vertices of the approximant (say all sites with a given $z$) and joining them by new edges yields a similar (rescaled) quasiperiodic tiling \citep{socolar89,duneau91}.  Consider,
for instance, sites with $z=4$. If we join them and scale the resulting
tiling by $s$, we recover the $z=6$ sites of the next approximant.
In the same fashion, the $z=3$ sites map into (part of) the $z=5$
sites. If we inflate an approximant twice, we recover all $z=8$ sites
of the second next approximant.


In Fig. \ref{fig:tiling_def}(b), we show the Fourier transform of
the sites with $z=3$ in the $N=8119$ approximant: $\left|\sum_{j}e^{-i\boldsymbol{r}_{j}\cdot\boldsymbol{k}}\right|^{2}/N$,  where $\boldsymbol{k}$ are the momentum and $\boldsymbol{r}_j$ the position of the site $j$ in the tiling. The brightest
Bragg peaks closest to the $\Gamma$ point are located at a distance
$\pi$ in units where the lattice spacing $a=1$. If we consider all
$N$ sites in the approximant, these peaks are instead located at
a distance $s\pi$.  Therefore,  focusing on the $z=3$ sites only,
we effectively rescale the lattice spacing by a factor $s$,  indicating that the effective neighbor distance of the quasicrystal is increasing. The inflation/deflation property thus guarantees the $8-$fold rotational symmetry of the approximant, even when considering only a specific subset of sites.

\subsection{Electronic states}

We model the electronic properties of this tiling considering a nearest-neighbor
tight-binding Hamiltonian 
\begin{equation}
H_{0}=-t\sum_{\left\langle ij\right\rangle ,\sigma}\left(c_{i\sigma}^{\dagger}c_{j\sigma}+c_{j\sigma}^{\dagger}c_{i\sigma}\right),\label{eq:tbh}
\end{equation}
where $c_{j\sigma}^{\dagger}\left(c_{j\sigma}\right)$ creates (destroys)
an electron of spin $\sigma$ at site $j$. $t$ is hopping amplitude
between nearest neighbors and is restricted to the sides of the rhombuses
and squares \citep{anu06}. We measure all energies in terms of $t$,
and the half-bandwidth is $\approx4t$, comparable to the one in the
square lattice, which is reasonable given that the average value of
$z$ is $4$. As we use open boundary conditions, we preserve the
particle-hole symmetry of this (bipartite) tiling and the spectrum is symmetric with respect to $\omega=0$,  allowing us to work solely with fillings $n\leq 1$.  In principle, by quadrupling the tilling size, it is possible to employ periodic boundary conditions while preserving the bipartness of the approximants \citep{jagannathan24}. In this work, we refrain from using this numerically more expensive approach.

We employ the KPM \citep{weisse06} to compute the spectral properties
of large approximants. This method was successfully applied in the previous
studies of disordered \citep{lee12,garcia15,yamada20} and quasicrystalline
\citep{wu22} systems, and is capable of resolving the low-energy features
of these systems. Within the KPM, the (rescaled) local density of
states is given by 
\begin{equation}
\rho_{ii}\left(\tilde{\omega}\right)=\frac{1}{\pi\sqrt{1-\tilde{\omega}^{2}}}\left[g_{0}\mu_{0}^{ii}+2\sum_{m=1}^{M}g_{m}\mu_{m}^{ii}T_{m}\left(\tilde{\omega}\right)\right],\label{eq:dos_kpm}
\end{equation}
where $T_{m}\left(\tilde{\omega}\right)$ is the $m-$th Chebyshev
polynomial, $M$ is the number of moments in the expansion, and $g_{m}$
is the Jackson Kernel \citep{weisse06}, introduced to dampen the
Gibbs oscillations due to a finite $M$. The expansion moments are
defined as $\mu_{m}^{ij}=\langle i|T_{m}\left(\tilde{H}\right)|j\rangle$,
where $\tilde{H}$ is the renormalized Hamiltonian with its spectrum
$\in\left[-1,1\right]$. The density of states can, in principle,
be computed as $\rho\left(\tilde{\omega}\right)=\sum_{i}\rho_{ii}\left(\tilde{\omega}\right)/N$,
which amounts to compute $\sum_{i}\langle i|T_{m}\left(\tilde{H}\right)|i\rangle=\text{Tr}\left[\mu_{m}^{ij}\right]$.
For large system sizes, the exact evaluation of traces is time-consuming,
and we resort to its stochastic estimation: $\text{Tr}\left[\mu_{m}^{ij}\right]=\sum_{r=1}^{R}\langle r|T_{m}\left(\tilde{H}\right)|r\rangle/R$,
where $|r\rangle$ is a random vector and $R$ is their number. The
advantage of this method is that we can use $R\ll N$ with its error
decreasing as $1/\sqrt{NR}$ \citep{weisse06}.

In Fig. \ref{fig:tiling_def}(c) we show $\rho\left(\omega\right)$
for $N=1607521$ sites and $R=10$. We can observe the key features
of the Ammann-Beenker tiling: (i) pseudogaps at energies corresponding
to the the filings $n=2/s^{4}$ and $n=2/s^{2}$, Fig. \ref{fig:tiling_def}(d);
(ii) the divergence of $\rho\left(\omega=0\right)$, due to localized
states; and (iii) the spiky nature of the $\rho\left(\omega\right)$
as a function of $\omega$ \citep{rieth95,anu06,janssen18}. The fact
that these key features are captured indicates that the KPM is
well suited to describe the non-trivial spectral dependence of the
electronic states in quasicrystal approximants,  up to their finite
size resolution.

\subsection{Magnetic model}

\begin{figure}[t]
\centering{}\includegraphics[width=1\columnwidth]{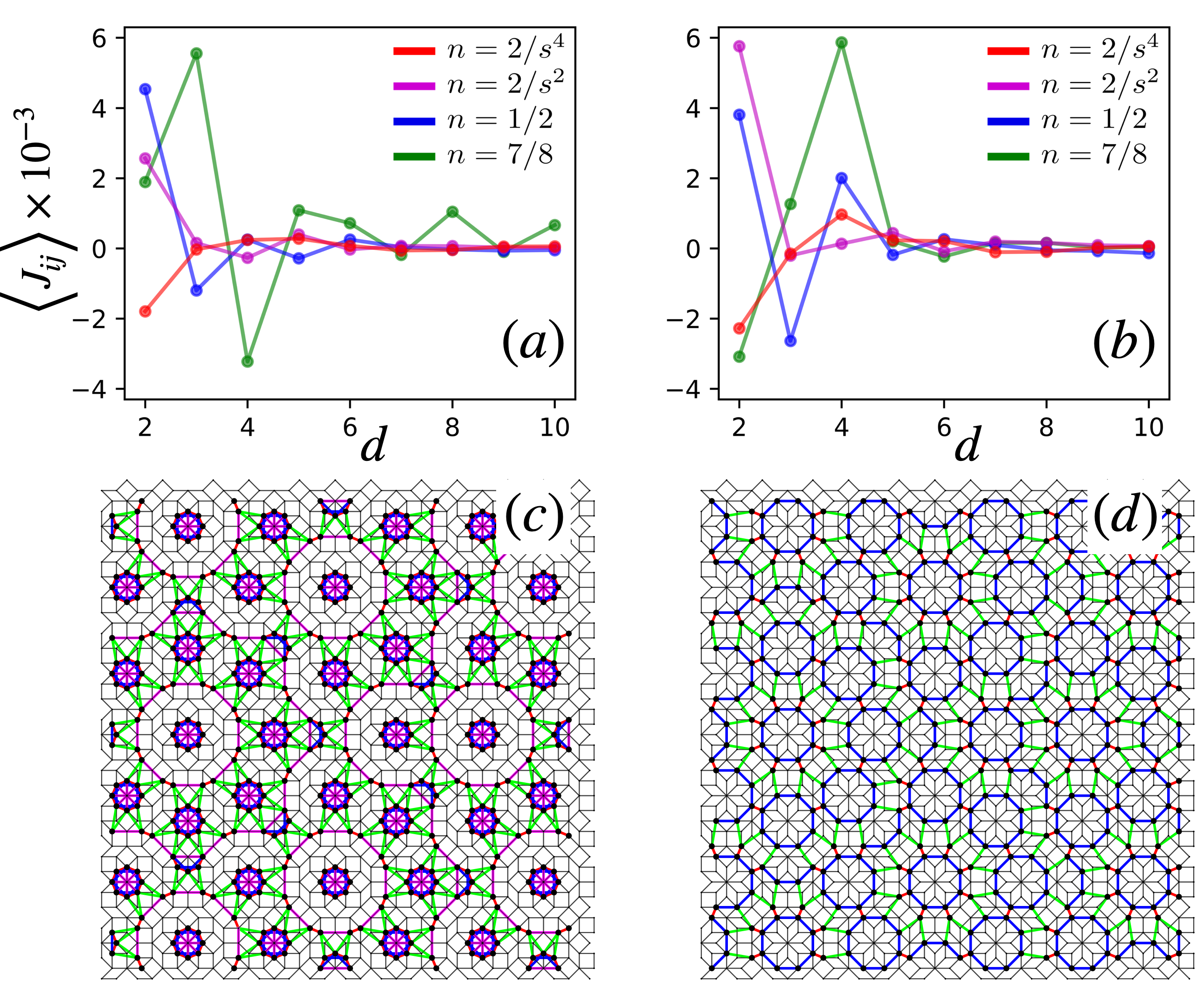}\caption{\label{fig:RKKY} Average value of the RKKY coupling in Eq. \eqref{eq:j_ij},
for several fillings $n$ and for sites with (a) $z=3$ and (b) $z=4$. We considered
$N=8119$, $J_{0}=t=1$,  and $d$ is the Manhattan distance
$d=\left|\Delta x\right|+\left|\Delta y\right|$, where $\Delta x \left(\Delta y\right)$
is the difference in the $x\left(y\right)$ coordinates of the sites
$i$ and $j$.  Minimal set of exchange couplings for the
$N=1393$ tiling. (c) $z=3$. Four magnetic exchange couplings: $J_{1}$
(red), $J_{2}$ (blue), $J_{3}$ (green) and $J_{4}$ (magenta). (d)
$z=4$. Three couplings: $J_{1}$ (red), $J_{2}$ (blue), and $J_{3}$
(green).}
\end{figure}

Our main motivation is to describe the fate of local magnetic moments in contact with a quasicrystalline electronic bath. Because quasicrystals display a distribution of local environments, we consider that the magnetic moments are not randomly placed but instead occupy all sites with a given $z$. We model these moments as a spin-$S$ and couple them to the local spin density of the conduction electrons via an exchange coupling $J_0$. To generate a spin-only model, we integrate out the conduction electron and write
\begin{equation}
H=\frac{1}{2}\sum_{i\neq j}J_{ij}S_{i}S_{j},\label{eq:ising}
\end{equation}
where $S_{i}=\pm1$ and the interaction between spins $J_{ij}$ is
\citep{lee12} 
\begin{equation}
J_{ij}=-2J_{0}^{2}\int_{E_{F}}^{D}d\omega\int_{-D}^{E_{F}}d\omega^{\prime}\frac{\rho_{ij}\left(\omega\right)\rho_{ji}\left(\omega^{\prime}\right)}{\omega-\omega^{\prime}},\label{eq:j_ij}
\end{equation}
where $D$ is the electronic half-bandwidth, $E_{F}$ is the Fermi
energy, and $\rho_{ij}\left(\omega\right)$ is given in Eq. \eqref{eq:dos_kpm}.
We consider $T=0$ and $M\sim10^{3}$ KPM moments to evaluate $J_{ij}$.
A few comments are in place: (i) Instead of considering the Heisenberg model, we study the simpler Ising model. This is motivated by the studies of spin glasses \citep{fischer93}, where the key aspects of the local moment freezing are already captured by Eq. ~\ref{eq:ising}; (ii) We include the effects of the conduction electrons at second order in $J_0$ as we expect Eq. ~\ref{eq:ising} to qualitatively describe the ordering of local moments in a quasicrystalline metallic environment.  This is physically motivated by the observation that,  although the impurities alter the electronic states of the bath,  as long as the conduction electrons remain metallic,  the key features of the effective magnetic couplings are captured perturbatively.

We show the average value of the coupling $J_{ij}$ for a given local
environment in Figs.  ~\ref{fig:RKKY}(a) and ~\ref{fig:RKKY}(b). As discussed previously
in Refs. \citep{thiem15a,thiem15b}, $J_{ij}$ has a complex spatial
and filing dependence. To obtain the full set of $J_{ij}$ required
for the MC simulations of Eq. \eqref{eq:ising} is a challenging task,
and we are restricted to $N\le8119$ approximants, even within KPM.
This hampers a conclusive finite-size scaling study of putative ordered
states. Moreover, $J_{ij}$ might display appreciable values for distant
sites \citep{jeon23}, further complicating a finite-size study.

To overcome these challenges, we employ a simpler model, following
Ref. \citep{duneau91}. Given a subset of sites, say sites with $z=3$
or $z=4$, we consider a minimal set of exchange couplings generating
unbounded clusters at $T=0$ and therefore allowing for long-range
order, Figs.  ~\ref{fig:RKKY}(c) and ~\ref{fig:RKKY}(d). For the $z=3$ environment,
we implement this set with an exchange coupling $J_{1}$ across the
short diagonal of the rhombus, a coupling $J_{2}$ across two neighboring
rhombi, an exchange $J_{3}$ across neighboring square and rhombus,
and a coupling $J_{4}$ across two squares, or rhombi, edges. The
pattern formed by this choice of couplings is displayed in Fig. \ref{fig:RKKY}(c),  where we see it percolates the lattice but still leaves a fraction
of the sites surrounding the $z=8$ sites weakly coupled to the bulk.

For $z=4$, the set of couplings is displayed in Fig. ~\ref{fig:RKKY}(d).  Here, we consider the $J_{1}$ coupling across the short diagonal
of the rhombus, a coupling $J_{2}$ across the square diagonal, and
the $J_{3}$ coupling across neighboring squares and rhombi. There
are no large clusters of disconnected structures for this minimal
choice of couplings. From now on,  we set $J_{1}$ as our energy scale. 

\begin{figure}[t]
\centering{}\includegraphics[width=1\columnwidth]{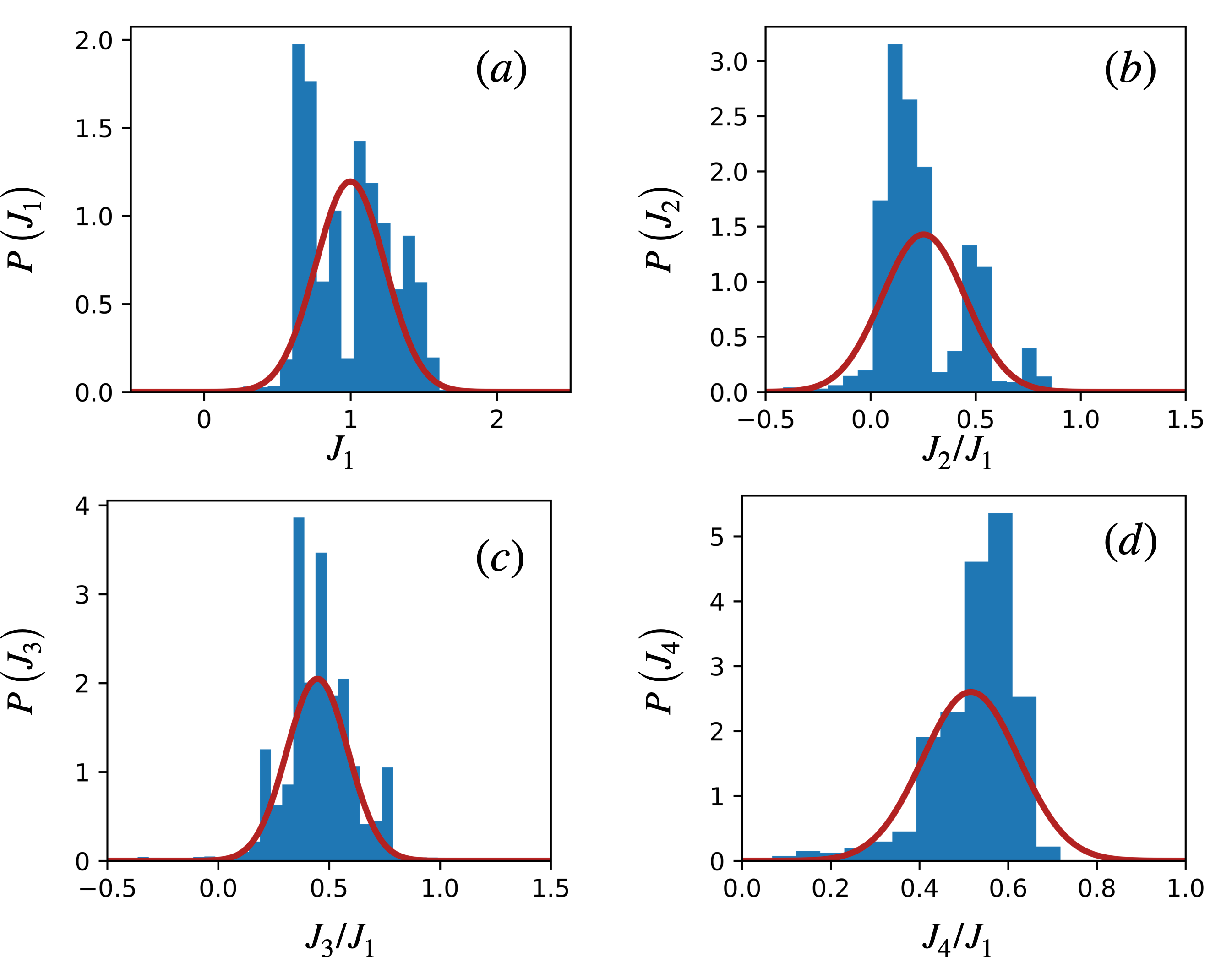}\caption{\label{fig:pj} Histogram of exchange couplings values for $z=3$
and $n=2/s^{2}$. We normalize the couplings with respect to $J_{1}$.
We superimpose a Gaussian fit with the average values of Tab. \ref{tab:z3}
and variances: (a) $P\left(J_{1}\right)$, $\sigma_{1}=0.33$; (b)
$P\left(J_{2}\right)$, $\sigma_{2}=0.28$; (c) $P\left(J_{3}\right)$,
$\sigma_{3}=0.20$; and (d) $P\left(J_{4}\right)$, $\sigma_{4}=0.15$.}
\end{figure}

We consider the value of these exchange couplings as the average value
of the corresponding RKKY interaction for the $N=47321$ approximant.
This approximant size is now amenable to KPM calculations because
we focus on a restricted subset of couplings.  In Fig. ~\ref{fig:pj}, we show the histogram of the minimal set of exchange couplings for $z=3$ and $n=2/s^2$. $P\left(J\right)$ is bounded and has a similar width for all neighbors, with its mean value capturing the correct energy scale of the problem. Therefore, the minimal model retains the chief ingredients to describe the order of local moments embedded in a metallic quasicrystal. In Appendix ~\ref{sec:further}, we show that the inclusion of extra couplings does not alter the conclusions presented in the main text.

\section{\label{sec:order}Magnetic order}

\begin{table}[t]
\begin{centering}
\begin{tabular}{|c|c|c|c|c|}
\hline 
$n$ & $J_{1}$ & $J_{2}$ & $J_{3}$ & $J_{4}$\tabularnewline
\hline 
\hline 
$2/s^{4}$ & $-1$ & $-0.501$ & $-0.163$ & $-0.321$\tabularnewline
\hline 
$2/s^{2}$ & $1$ & $0.255$ & $0.449$ & $0.518$\tabularnewline
\hline 
$1/2$ & $1$ & $-0.389$ & $-0.435$ & $0.042$\tabularnewline
\hline 
$7/8$ & $1$ & $0.666$ & $0.682$ & $0.559$\tabularnewline
\hline 
\end{tabular}
\par\end{centering}
\caption{\label{tab:z3}Exchange couplings $J_{i}$ for $z=3$ and different
fillings $n$. $\left|J_{1}\right|$ sets the scale.}
\end{table}

\begin{table}[t]
\begin{centering}
\begin{tabular}{|c|c|c|c|}
\hline 
$n$ & $J_{1}$ & $J_{2}$ & $J_{3}$\tabularnewline
\hline 
\hline 
$2/s^{4}$ & $-1$ & $-0.971$ & $-0.342$\tabularnewline
\hline 
$2/s^{2}$ & $1$ & $0.379$ & $-0.056$\tabularnewline
\hline 
$1/2$ & $-1$ & $3.242$ & $-3.701$\tabularnewline
\hline 
$7/8$ & $-1$ & $-0.831$ & $0.813$\tabularnewline
\hline 
\end{tabular}
\par\end{centering}
\caption{\label{tab:z4}Same as Tab. \ref{tab:z3} but now for $z=4$.}
\end{table}

We now discuss the magnetic states of Ising spins interacting with
the averaged RKKY couplings in Tabs. \ref{tab:z3} and \ref{tab:z4},
corresponding to four different fillings $n$ of the conduction electrons. To solve Eq. \eqref{eq:ising},
we employ classical MC simulations on approximants with $N=239,1393,8119,47321$
and open boundary conditions.  Within our scheme,  system size and aperiodicity are intertwined.  As we increase $N$, we increase the effective size of the approximant unit cell,  and the limit $N \to \infty$ corresponds to the quasicrystal.  We perform equilibrium MC simulations
combining single-site Metropolis updates with the parallel tempering
method \citep{hukushima96,newman99}. Usually, we perform $10^{5}$
Monte Carlo steps for thermalization and $5\times10^{5}$ Monte Carlo
steps for measurements.

To characterize long-ranged ordered phases, we compute the staggered
magnetization for a single Monte Carlo snapshot $\mathcal{M}_{s}=\sum_{i}\zeta_{i}S_{i}/N$,
with the Monte Carlo order parameter given by $m_{s}=\sqrt{\left\langle \mathcal{M}_{s}^{2}\right\rangle }$,
where $\left\langle \cdots\right\rangle $ denotes the Monte Carlo
average. The local phase $\zeta_{i}=\pm1$ describes the ordered ground
state spin configuration, and we will explain how we determine it shortly.
To capture the ordering temperature $T_{N}$, we employ the Binder
cumulant $B_{s}=3/2-\left\langle \mathcal{M}_{s}^{4}\right\rangle /2\left\langle \mathcal{M}_{s}^{2}\right\rangle ^{2}$,
normalized such that $B_{s}\to1$ in the ordered phase and $B_{s}\to0$
in the disordered phase. The crossing point of the curves $B_{s}\times T$
for different systems sizes determines $T_{N}$ \citep{newman99}.
To track possible spin freezing without long-range order, we also
compute the Edwards-Anderson order parameter \citep{fischer93,binder_young}:
$q_{i}=\left(1/\tau\right)\int_{0}^{\tau}S_{i}\left(0\right)S_{i}\left(\tau\right)$,
with $\tau$ the Monte Carlo simulation time, measured after the equilibration.
From this definition, we see that $q_{i}$ is the overlap of a spin
with itself at a later time. In a paramagnetic phase, $q\to0$ because
the spins fluctuate over time. If the spins are frozen in an ordered
state, we have both $q\neq0$ and $m_{s}\neq0$. For a spin-glass
state, $q\neq0$ and $m_{s}=0$. 

To calculate the local phases $\zeta_{i}$, corresponding to the spin
configuration in the ground state, we proceed as follows: (i) we perform
Monte Carlo runs at low temperatures and generate $10^{3}$ independent
spin configurations; (ii) we express Eq. \eqref{eq:ising} as $H=\sum_{i}S_{i}h_{i}/2$,
with the local (exchange) field defined as $h_{i}=\sum_{j}J_{ij}S_{j}$. We sequentially
sweep the lattice and iteratively anti-align a given spin with $h_{i}$
until the change in the spin configuration after the sweep is smaller than a given
tolerance value; (iii) out of the original $10^{3}$ states, we select
the spin configuration with the lowest energy as the global ground
state. This particular state defines the local phases $\zeta_{i}$ in our simulations for a given set of parameters \citep{thiem15a, thiem15b}.

The Ammann-Beenker tiling is bipartite, as illustrated by the particle
hole symmetric density of states in Fig. \ref{fig:tiling_def}(c).
If we place Ising spins on all approximant sites and allow them to
interacting with each other via a nearest-neighbor antiferromagnetic
coupling $J_{1}$, we get a Néel ground state \citep{ledue95, repetowicz99, wessel03}
below $T_{N}/J_{1}=2.386\left(2\right)$. Therefore, a dense quasiperiodic
lattice of local moments can sustain regular long-range order \citep{cabrera-baez19}.

\subsection{$z=3$ sites}

\begin{figure}[t]
\centering{}\includegraphics[width=1\columnwidth]{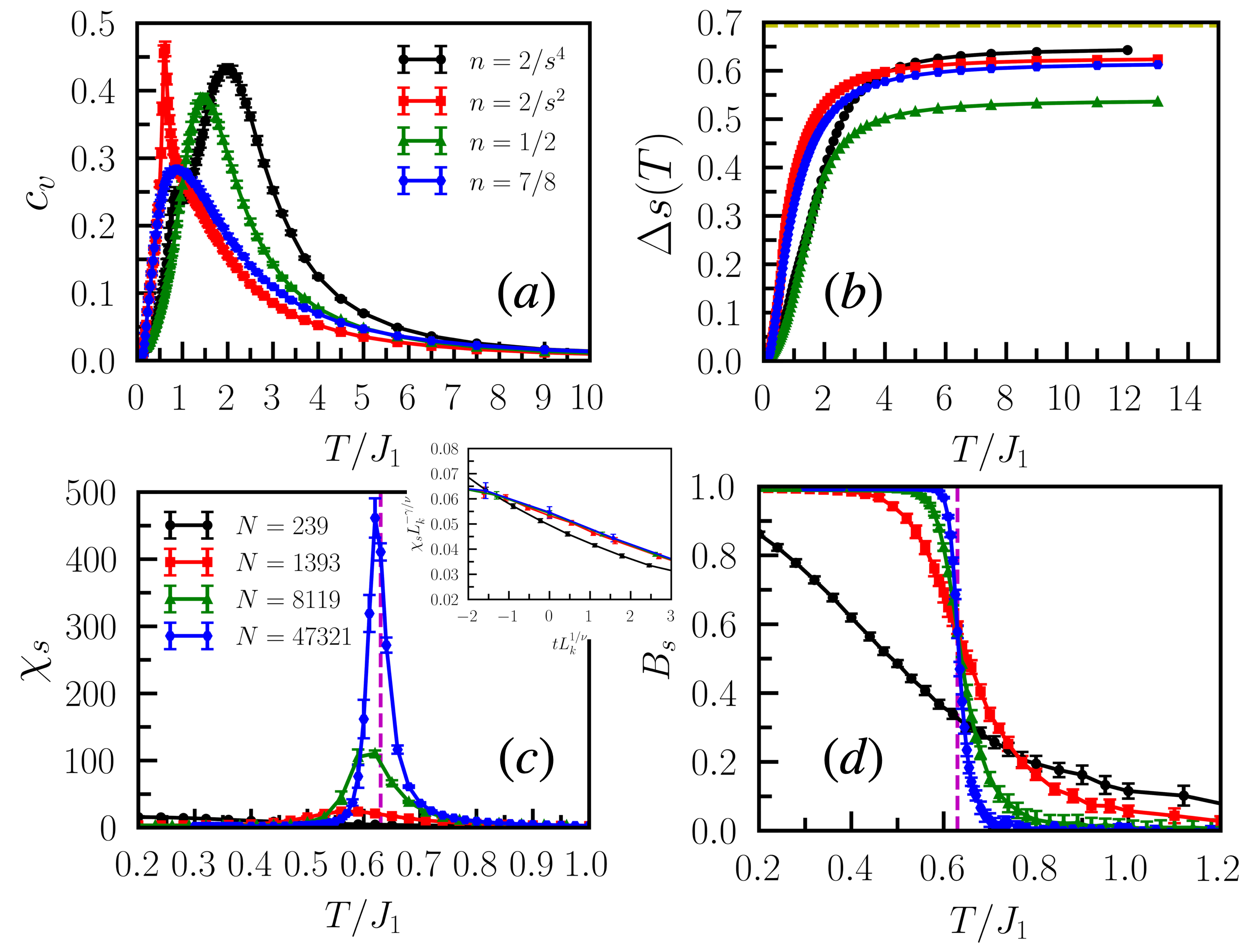}\caption{\label{fig:z3_MC_thermo}Monte Carlo results for local moments placed
at sites with $z=3$ local environments in the Ammann-Beenker tiling.
(a) Specific heat $c_{v}$ as a function of temperature $T$ for four
different conduction electron fillings $n$ and $N=8119$. (b) Residual
entropy $\Delta s\left(T\right)$, per site, as a function of $T$
for $N=8119$. The dashed line gives $s\left(T\to\infty\right)=\log{2}$.
The curve labels are the same as in (a). (c) Susceptibility associated
with the staggered magnetization $\chi_{s}$ as a function of $T$
for $n=2/s^{2}$ and different $N$. The vertical dashed line marks
the position of the ordering temperature $T_{N}$.  Inset: Finite scaling of the data. Here, $t=\left(T-T_N\right)/T_N$ is the reduced temperature, $L_k$ is the linear size of the approximant (see text), and we consider the critical exponents corresponding to the two-dimensional Ising universality class: $\gamma=7/4$ and $\nu=1$.
(d) Binder cumulant $B_{s}$ of the order parameter as a function
of $T$ for $n=2/s^{2}$. The dashed line marks the crossing of the
curves for $N\ge1393$ and determines $T_{N}=0.63\left(1\right)J_{1}$.
The curve colors are the same as in (c).}
\end{figure}

We start discussing the situation where the magnetic impurities are
placed in all sites with $z=3$, corresponding to a fraction of $f_{3}=1/s=0.414$
of the $N$ sites in an approximant, see Fig. \ref{fig:tiling_def}(a).
The corresponding magnetic coupling structure is in Fig. \ref{fig:RKKY}
(c), with the corresponding $J_{i}$ values in Tab. \ref{tab:z3}.

The specific heat displays a peak for all electronic fillings $n$
considered in this study, Fig. \ref{fig:z3_MC_thermo}(a). The peak
position depends on $n$ since the magnetic couplings are a function
of $n$. We estimate the residual entropy, per spin, as $\Delta s\left(T\right)=\int_{0}^{T}\left(c_{V}/T\right)dT=s\left(T\right)-s_{0}$.
For an ordinary ordered system, we expect $s_{0}\to0$ and $s\left(T\to\infty\right)\to\log2$.
Fig. \ref{fig:z3_MC_thermo}(b) shows that $\Delta s$ never saturates
to $\log2$, indicating a residual ground state entropy, per site,
$s_{0}\neq0$. If we ascribe this residual entropy to the weakly coupled
clusters of spins with $z=3$ surrounding the $z=8$ sites \citep{thiem15a,thiem15b},
we can estimate the ground state degeneracy, per spin, as $g_{0}\approx\left(1/4\right)\times8f_{8}/f_{3}=2/s^{3}$.
The factor of $1/4$ comes from the fact that about $1/4$ of these
clusters are weakly coupled, Fig. \ref{fig:RKKY}(c). This amounts
to $\Delta s\left(T\to\infty\right)=\log2\left(1-2/s^{3}\right)\approx0.59$,
which corresponds roughly to the values presented in Fig. \ref{fig:z3_MC_thermo}(b)
for all fillings but $n=1/2$. This last case is special because here
$J_{4}\approx0$,  enhancing the ground-state degeneracy.

\begin{figure}[t]
\centering{}\includegraphics[width=1\columnwidth]{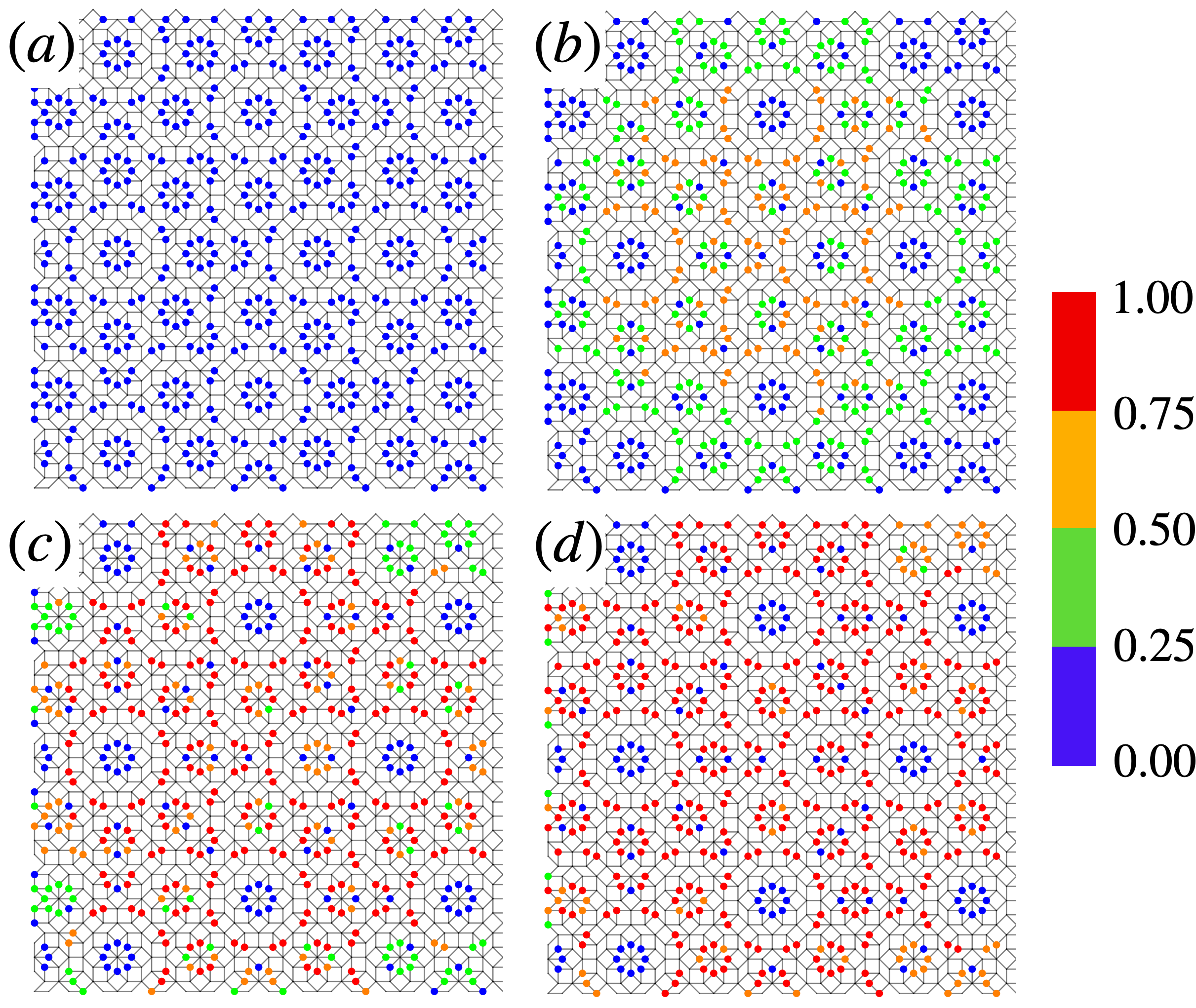}\caption{\label{fig:qfluc}Spatial distribution of the Edwards-Anderson
order parameter $q_{i}$, at different temperatures, for local moments
placed at sites with $z=3$ local environments in the Ammann-Beenker
tiling. (a) $T=0.60J_{1}$; (b) $T=0.50J_{1}$; (c) $T=0.30J_{1}$,
(d) $T=0.20J_{1}$. Here we considered $n=2/s^{2}$, $N=1393$, and
$T_{N}=0.63\left(1\right)J_{1}$ for this parameter set. The color defines the value of $q_i$.}
\end{figure}

To corroborate this picture, we study the spatial structure of the spin overlap $q_i$ for a given temperature $T$, Fig. ~\ref{fig:qfluc}. We observe that for $T>T_N$, $q_i \approx 0$ for all sites. As the temperature decreases, $q_i$ increases, suggesting an ordered state. However, even at low-$T$, roughly a quarter of the $z=3$ sites surrounding the $z=8$ sites show $q_i \approx 0$, indicating that they indeed fluctuate despite the bulk of the spins being ordered.

Fig. \ref{fig:z3_MC_thermo}(c) shows the susceptibility associated
with the staggered order parameter for the electronic filling $n=2/s^{2}$.
Even though $s_{0}\neq0$, the system orders, as indicated by the
enhancement of the peak in $\chi_{s}$ as we increase the system size. 
The Binder cumulant captures this critical behavior in Fig. \ref{fig:z3_MC_thermo}(d),
with the curves for different $N$ crossing at the ordering temperature
$T_{N}=0.63\left(1\right)J_{1}$. The curve for $N=239$ contains
too few spins and suffers from strong finite-size corrections, crossing
the other curves at different points. The Curie-Weiss temperature
for this parameter set is $\theta_{CW}\approx-3J_{1}$, implying the
frustration pushes the ordered temperature down by a factor of $5$
compared to this scale.  With the value of $T_N$,  we check that $\chi_s$ follows the usual finite-size scaling with the critical exponents for the two-dimensional Ising universality class,  inset of Fig. ~\ref{fig:z3_MC_thermo}(c) ~\citep{sorensen91, luck93, ledue95, repetowicz99, gallone24}.  We then conclude that the magnetic transition survives for $N \to \infty$,  ruling out, for instance, the existence of an intrinsic spin-glass phase. 

Most magnetic moments develop a long-range order while a small fraction of the spins keep fluctuating
at low temperatures. This is similar to partially ordered states identified
in frustrated magnets. It happens because the ordered spins do not
exert any mean field on the remaining disordered ones, and thermal
fluctuations are unable to lift this degeneracy \citep{diep97,wills02,javanparast15,gonzalez19,seifert19,partial_kag}.

\begin{figure}[t]
\centering{}\includegraphics[width=1\columnwidth]{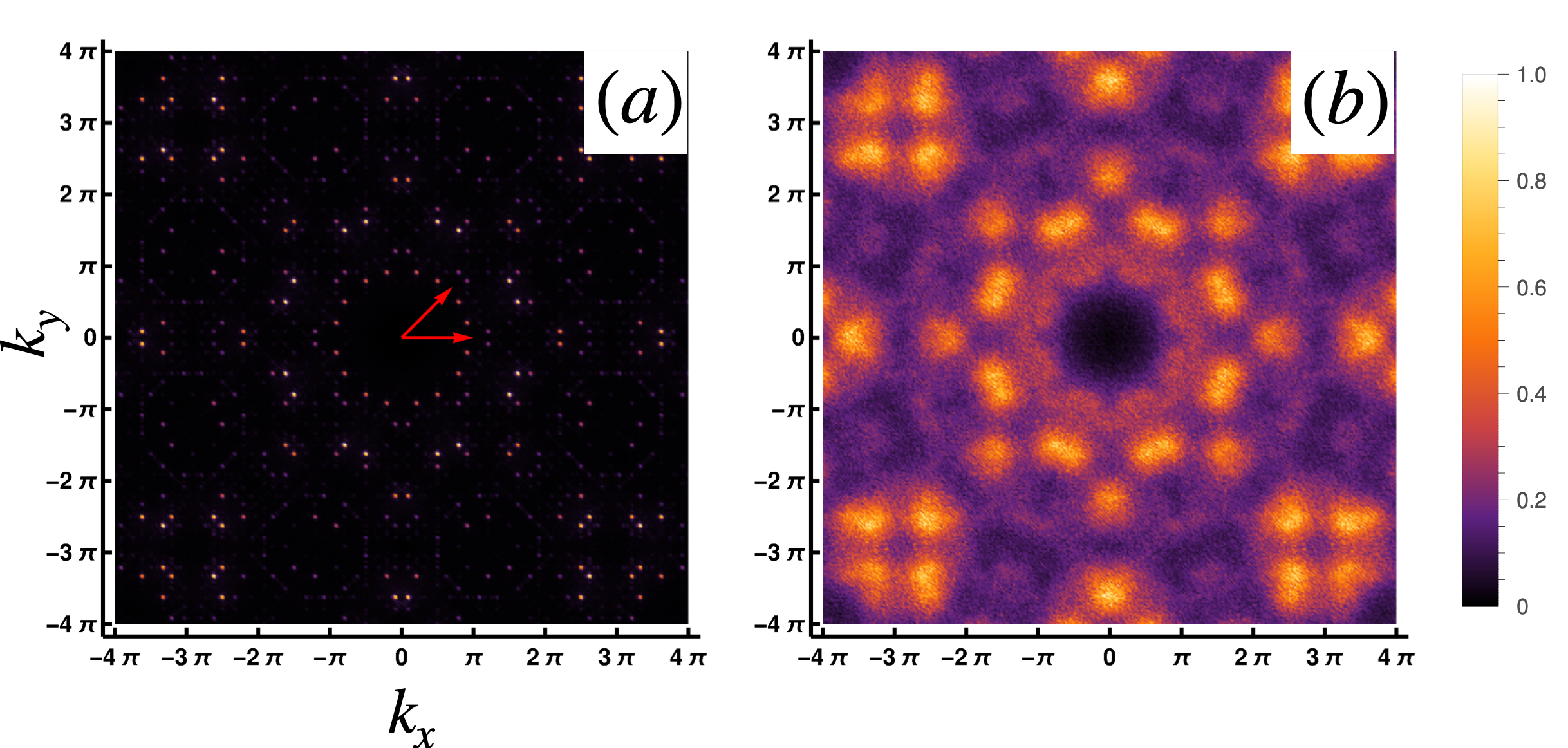}\caption{\label{fig:static_structure}Static spin structure factor for local
moments placed at sites with local environments $z=3$. We consider
$N=8119$ and $n=2/s^{2}$. (a) $T=0.5J_{1} <  T_N$. (b) $T = 1.0J_{1} > T_N$.
The red arrows are the same as in Fig. \ref{fig:tiling_def}(b). The
color code in each figure is independently normalized.}
\end{figure}

\begin{figure}[t]
\centering{}\includegraphics[width=1\columnwidth]{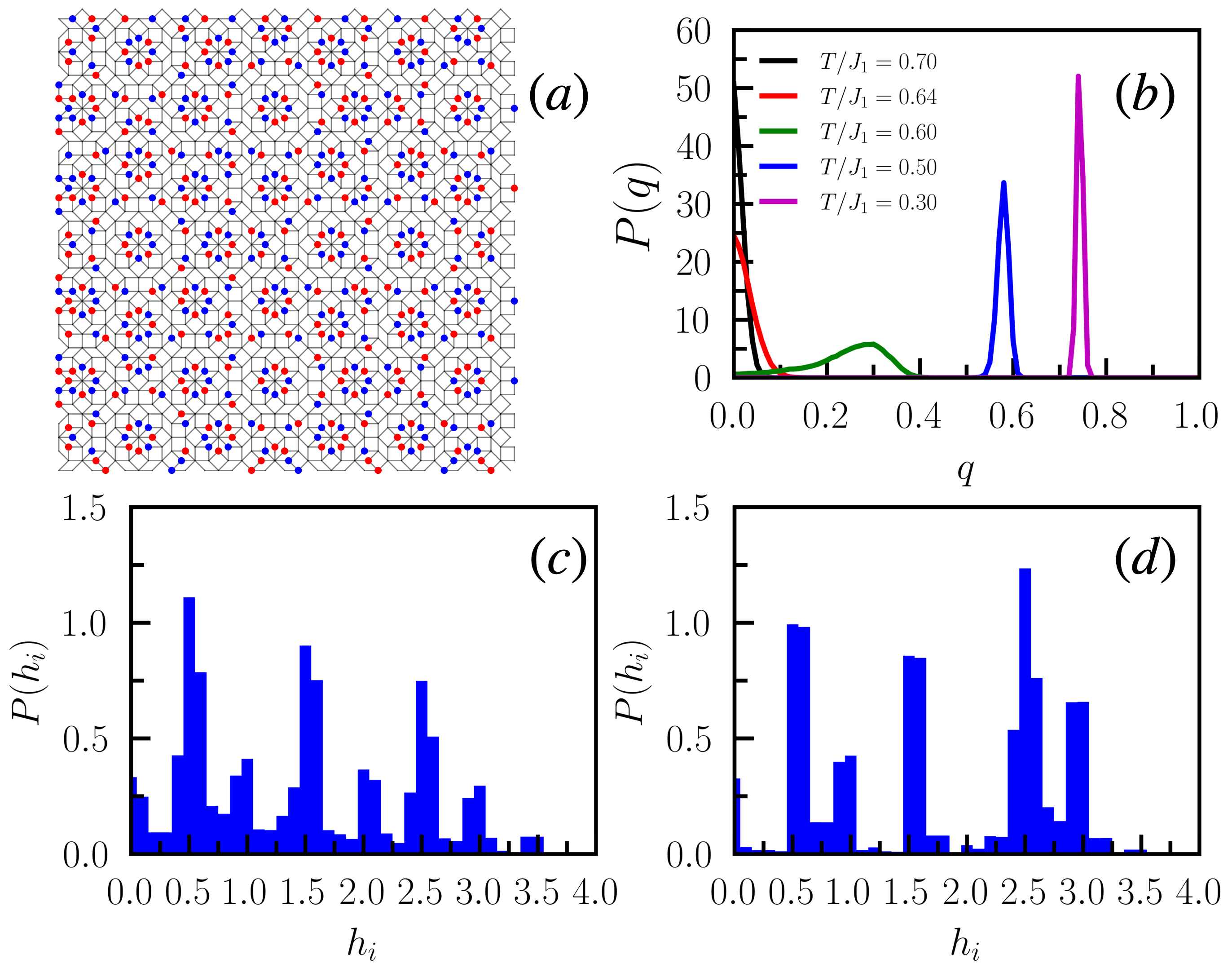}\caption{\label{fig:orderz3}Characterization of the ordered states for local
moments placed at the $z=3$ sites and $n=2/s^{2}$. (a) Ground state
configuration for $N=1393$. The different colors indicate opposite
spin orientations defining the local phases $\zeta_i=\pm1$ (see text).  (b) Histogram of the Edwards-Anderson order parameter
$P\left(q\right)$ for different temperatures $T$. Histogram of the
local exchange field, $P\left(h_{i}\right)$: (c) $T=J_{1}>T_{N}$
and (d) $T=0.3J_{1}<T_{N}$. $P\left(q\right)$ and $P\left(h_{i}\right)$
are both symmetric with respect to zero due to time reversal symmetry.
Therefore, we consider only positive values of $q$ and $h_{i}$.}
\end{figure}

We investigate the resulting ordered states via the static spin structure
factor,  $\left<\left|\sum_{j} S_j e^{-i\boldsymbol{r}_{j}\cdot\boldsymbol{k}}\right|^{2}\right>/N$,  with representative results in Fig. \ref{fig:static_structure}.
Above the ordering temperature, the structure factor displays the
$8$-fold symmetry inherited from the tiling with the Bragg-like peaks
broadened by the temperature. For $T<T_{N}$ the brightest peaks split
into two, and their position does not coincide with the structural
peaks in Fig. \ref{fig:tiling_def}(b), indicating the emergence of
a complex antiferromagnetic arrangement of the local moments. The
existence of well-defined bright peaks at low$-T$ also confirms that
the fluctuating spins occupy well-defined positions in the lattice.
A typical ground state with local moments displaying long-range antiferromagnetic
order is shown in Fig. \ref{fig:orderz3}(a). This state defines the local phases $\zeta_i=\pm1$.

\begin{figure}[t]
\centering{}\includegraphics[width=1\columnwidth]{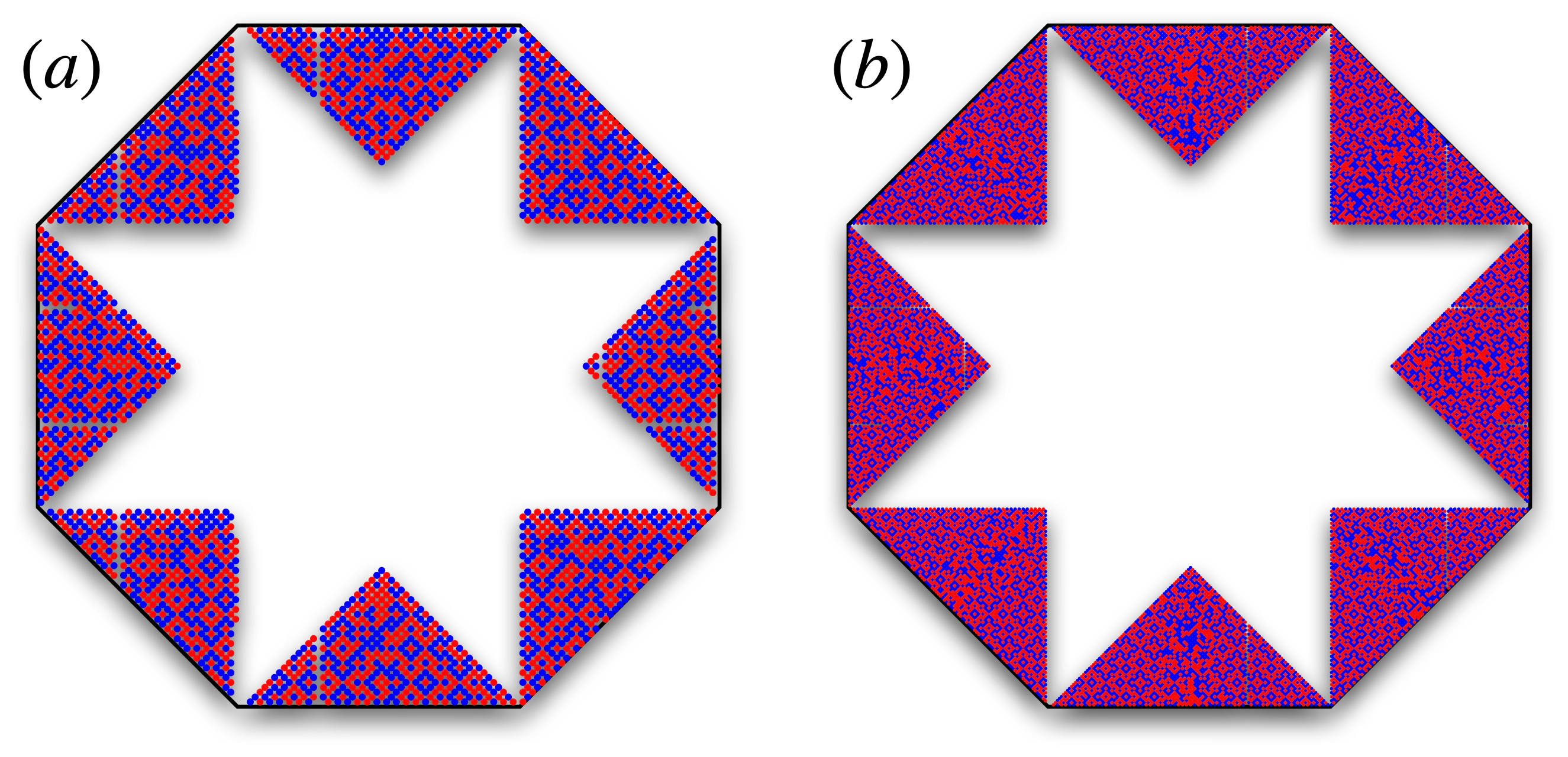}\caption{\label{fig:orderz3_perp}Characterization of the ordered states for local
moments placed at the $z=3$ sites and $n=2/s^{2}$ in the perpendicular space for (a) $N=8119$ and (b) $N=47321$.  The different colors indicate opposite spin orientations.  We exclude the sites siting at the open boundaries.}
\end{figure}

An useful way of studying the behavior of local quantities in a quasicrystal is to consider the so-called perpendicular space \cite{sakai19,  ghadimi20}.  Within the cut and project method,  Ref. \citep{deneau89},   we consider a parent $4$-dimensional cubic lattice which has  two  orthogonal  bidimensional planes,  called  the  physical  and the perpendicular planes,  both symmetric under $8$-fold rotations.  The approximant tilings are in the physical space.  The perpendicular space organizes the sites according to their local environment.  Because in our work we consider a subset of $z$,  we select the corresponding regions in this space.  A plot of the ground state magnetization for two system sizes, with $z = 3$ sites and $n = 2/s^2$ is shown in Fig. \ref{fig:orderz3_perp}.  No immediate pattern, or significant size dependence,  is visible inside each region indicating that the local value of the magnetization depends on a complex balance between the tiling geometry and the frustration induced by the magnetic couplings.  Therefore, we focus on analyzing the results in real or physical space. 

We start by investigating the distribution
of the Edwards-Anderson order parameter, $P\left(q\right)$, Fig.
\ref{fig:orderz3}(b). This histogram is constructed considering the values of $q=\sum_{i}q_{i}/N$
for different configurations, as those of Fig. \ref{fig:qfluc},  at  several Monte Carlo times $\tau$.  For $T>T_{N}$, the spins are
fluctuating and $P\left(q\right)$ is peaked at $q=0$, see also Fig.
\ref{fig:qfluc}(a). For $T<T_{N}$, $P\left(q\right)$ peaks at  $q \neq 0$, indicating the gradual freezing of the spins into
the ordered state; see also Figs. \ref{fig:qfluc}(b)-(d). Because
a fraction of the spins continue fluctuating to very low temperatures,
$P\left(q\right)$ peaks at values $q<1$ even at the lowest $T$,
thus accounting for a non-vanishing residual entropy $s_{0}$. In
our simplified model, we always detect a single peak in $P\left(q\right)$,
signaling a unique ground state up to symmetry-allowed operations
\citep{thiem15a,thiem15b}.

Alternatively,  we investigate the ordered state via the distribution
of local, or mean, fields $h_{i}=\sum_{j}J_{ij}S_{j}$, which corresponds
to the density of states of Eq. \eqref{eq:ising}. The histograms
$P\left(h_{i}\right)$ in Figs. \ref{fig:orderz3}(c) and (d) display
many peaks, indicating an intricate local field distribution due to
the combination of quasiperiodicity and a complex ordered state,  Fig. ~\ref{fig:orderz3}(a). Importantly, even for $T<T_{N}$, we observe
$P\left(h_{i}\to0\right)\neq0$, corroborating the presence of fluctuating
spins inside the ordered phase.

\begin{figure}[t]
\centering{}\includegraphics[width=1\columnwidth]{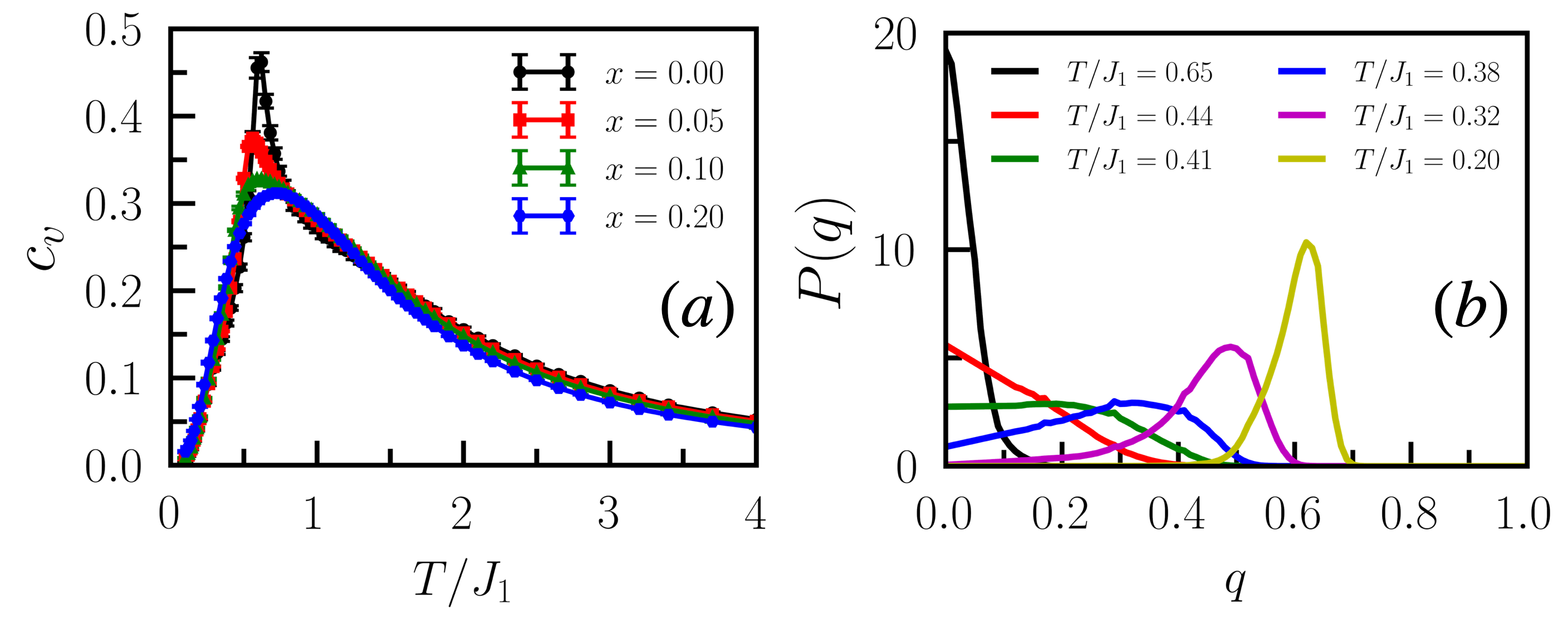}\caption{\label{fig:orderz3_dilution}Monte Carlo results for local moments
placed at $z=3$ local environments, $N=8119$, and $n=2/s^{2}$.
(a) $c_{v}$ as a function of $T$ for different values of dilution
$x$. (b) $P\left(q\right)$ for $x=0.10$ and different $T$ values.}
\end{figure}

Taken together, these facts point toward a fragile, long-range, ordered
state. Because $P\left(h_{i}\to0\right)\neq0$, perturbations will
couple directly to the loose spins, and those at small but finite
$h_{i}$, profoundly affecting the ordered state even at weak coupling
\citep{shiino21}. To illustrate this point, we dilute a fraction
$x$ of the spins in a given sample. We also average our results over
different vacancy configurations in these simulations, typically $100$.
In Fig. \ref{fig:orderz3_dilution}(a), we observe that the specific
heat peak is rounded and becomes broader as we increase the vacancy
concentration, a typical signature of a glassy state \citep{fischer93}.  

A peak in the specific heat marks the location of a relevant microscopic energy scale in the problem. In a spin glass, we have a distribution of energy scales due to the quenched disorder, and thus,  we expect a broad peak in the specific heat.  Moreover,  as we no longer have a transition from a paramagnetic state to an ordered state,  with the accompanying substantial release of entropy,  the peak in the specific heat becomes rounded.
We confirm the presence of a glass-like nature of the magnetic state
by investigating $P\left(q\right)$. At not-too-low temperatures,
it peaks at $q\neq0$ and a tail extending to a finite value at $q=0$
\citep{young83}. As we further reduce the temperature, the correlation
length becomes comparable to the system size, causing $P\left(q\right)$
to display a broad peak at $q<1$ \citep{bray84}.

\subsection{$z=4$ sites}

We present sample results for placing local moments at sites with
$z=4$ neighbors. They correspond to a fraction $f_{4}=2/s^{2}=0.343$
of all sites in a given approximant. Sites with $z=3$ are particular
because they display weakly connected islands around the $z=8$ sites.
Sites with $z=4$ form a more connected network of interacting spins,
Fig. \ref{fig:RKKY}(d). The corresponding values of the magnetic
couplings are presented in Tab. \ref{tab:z4}.

\begin{figure}[t]
\centering{}\includegraphics[width=1\columnwidth]{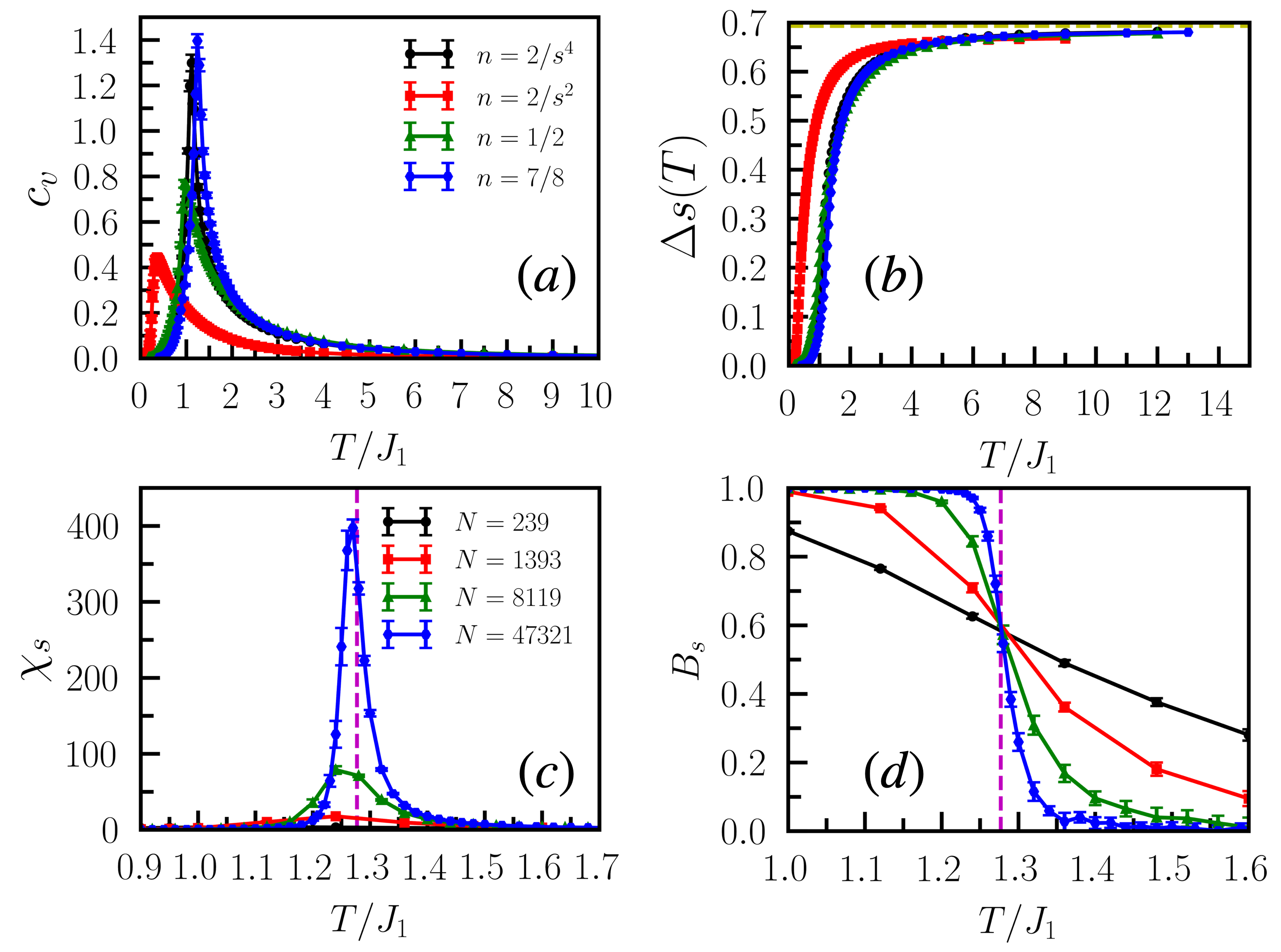}\caption{\label{fig:z4_MC_thermo}Monte Carlo results for local moments placed
at $z=4$ local environments. (a) Specific heat $c_{v}$ as a function
of $T$ for four $n$ and $N=8119$. (b) Residual entropy $\Delta s\left(T\right)$,
per site, as a function of $T$ for $N=8119$. The dashed line gives
$s\left(T\to\infty\right)=\log{2}$. The curve labels are the same
as in (a). (c) $\chi_{s}$ as a function of $T$ for $n=7/8$. The
vertical dashed line marks the ordering temperature's position $T_{N}$.
(d) Binder cumulant $B_{s}$ as a function of $T$ for $n=7/8$. The
dashed line marks the crossing of the curves and determines $T_{N}=1.277\left(5\right)J_{1}$.
The curve colors are the same as in (c).}
\end{figure}

Fig. \ref{fig:z4_MC_thermo}(a) shows the specific heat as a function
of the temperature. Compared to Fig. \ref{fig:z3_MC_thermo}(a), the
peaks are more pronounced, indicating more robust ordered states.
This is confirmed in Fig. \ref{fig:z4_MC_thermo}(b), where we see
that the residual entropy $s_{0}$ is tiny. Using MC simulations,
it is hard to pinpoint the precise value of $s_{0}$ when it is small
\citep{colbois22}, and we take $s_{0}\approx0$.

A representative study of the ordering is shown in Figs. \ref{fig:z4_MC_thermo}(c)
and (d) for $n=7/8$. The staggered susceptibility displays a well-defined
peak close to the ordering temperature $T_{N}/J_{1}=1.277\left(5\right)$,
which we extract via the crossing of the Binder cumulant. Here, even
the curve from the smallest system size crosses all the others at
$T_{N}$, indicating milder finite size effects than the $z=3$ case.
The Curie-Weiss temperature for this parameter set is $\theta_{CW}\approx2.5J_{1}$,
implying the effects of frustration are not so severe despite the
lack of periodicity and competing magnetic interactions.

\begin{figure}[t]
\centering{}\includegraphics[width=1\columnwidth]{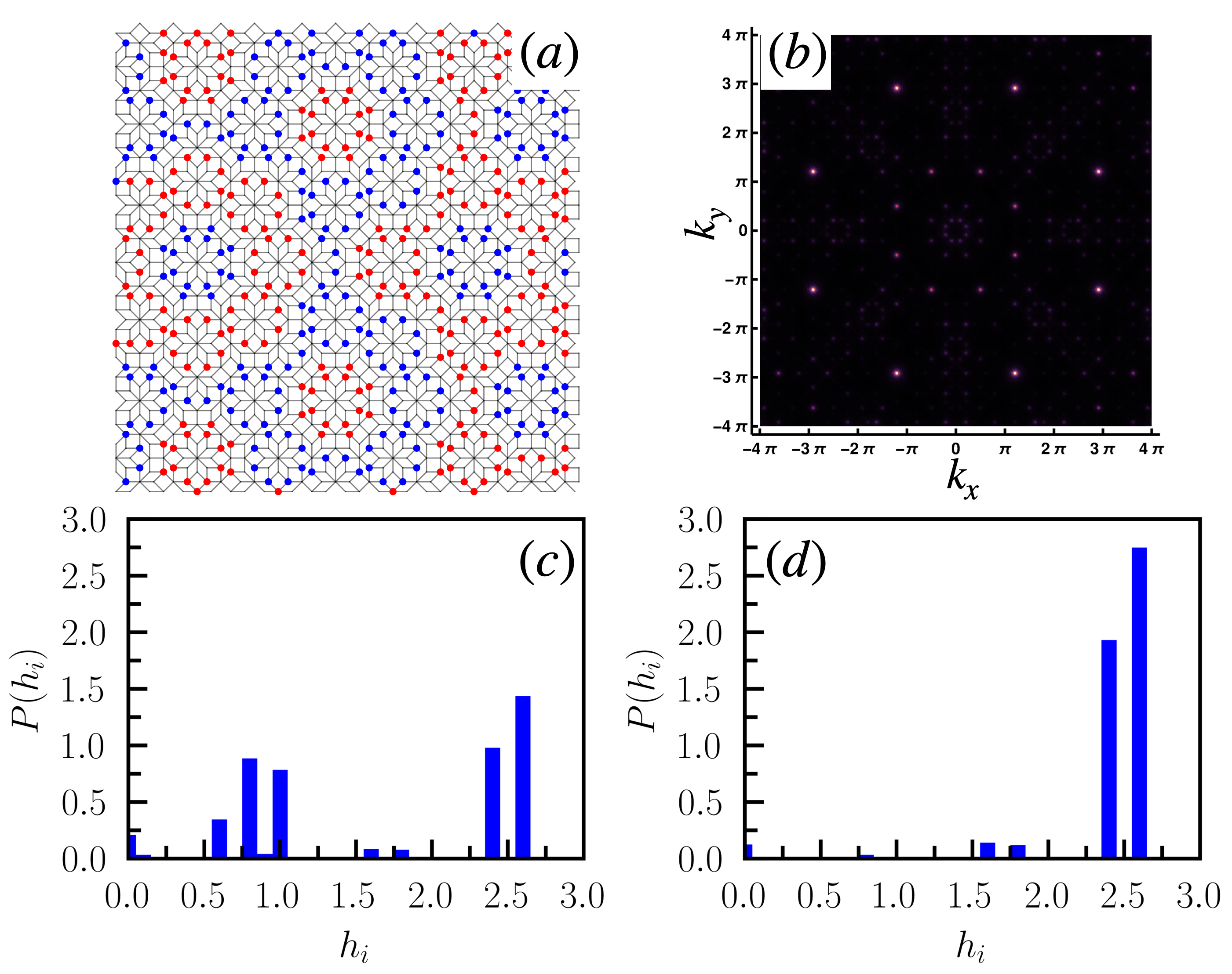}\caption{\label{fig:orderz4}Characterization of the ordered states for local
moments place at the $z=4$ sites and $n=7/8$. (a) Ground state configuration
for $N=1393$. The different colors indicate opposite spin orientations.
(b) Static spin structure factor for $N=8119$ and $T/J_{1}=1.2$.
The color scale is the same used in Fig. \ref{fig:static_structure}.
Histogram of the local exchange field, $P\left(h_{i}\right)$: (c)
$T=1.4J_{1}$ and (d) $T=0.5J_{1}$.}
\end{figure}

The ground state for $n=7/8$ is shown in Fig. \ref{fig:orderz4}(a).
It displays islands of spins up (down) surrounded by islands of spin
down (up), with the spins across the rhombus short diagonal parallel
to each other. The structure factor at finite temperature is shown
in Fig. \ref{fig:orderz4}(b), with the sharpest Bragg-like peaks
displaying the $8$-fold symmetry and no splitting, reflecting the
ordered state's local ferromagnetic structure. To gauge the robustness
of the ordered state for $n=7/8$, we compute $P\left(h_{i}\right)$,
Fig. \ref{fig:orderz4}(c) and (d). We observe that the number of
peaks and the relative contribution of the peak at $h_{i}\to0$ are
smaller than those in Fig. \ref{fig:orderz3}. This is consistent with a more
robust and simpler ordered pattern,  signaled by the absence of weakly coupled spin-cluster and $s_{0}\approx0$.  Because the system is frustrated,
the disorder will eventually melt the ordered state \citep{cantarino19,michel21,ye22},
but one expects a less pronounced effect in comparison to the $z=3$
case.

\section{\label{sec:penrose}Penrose lattice}

\begin{figure}[t]
\centering{}\includegraphics[width=1\columnwidth]{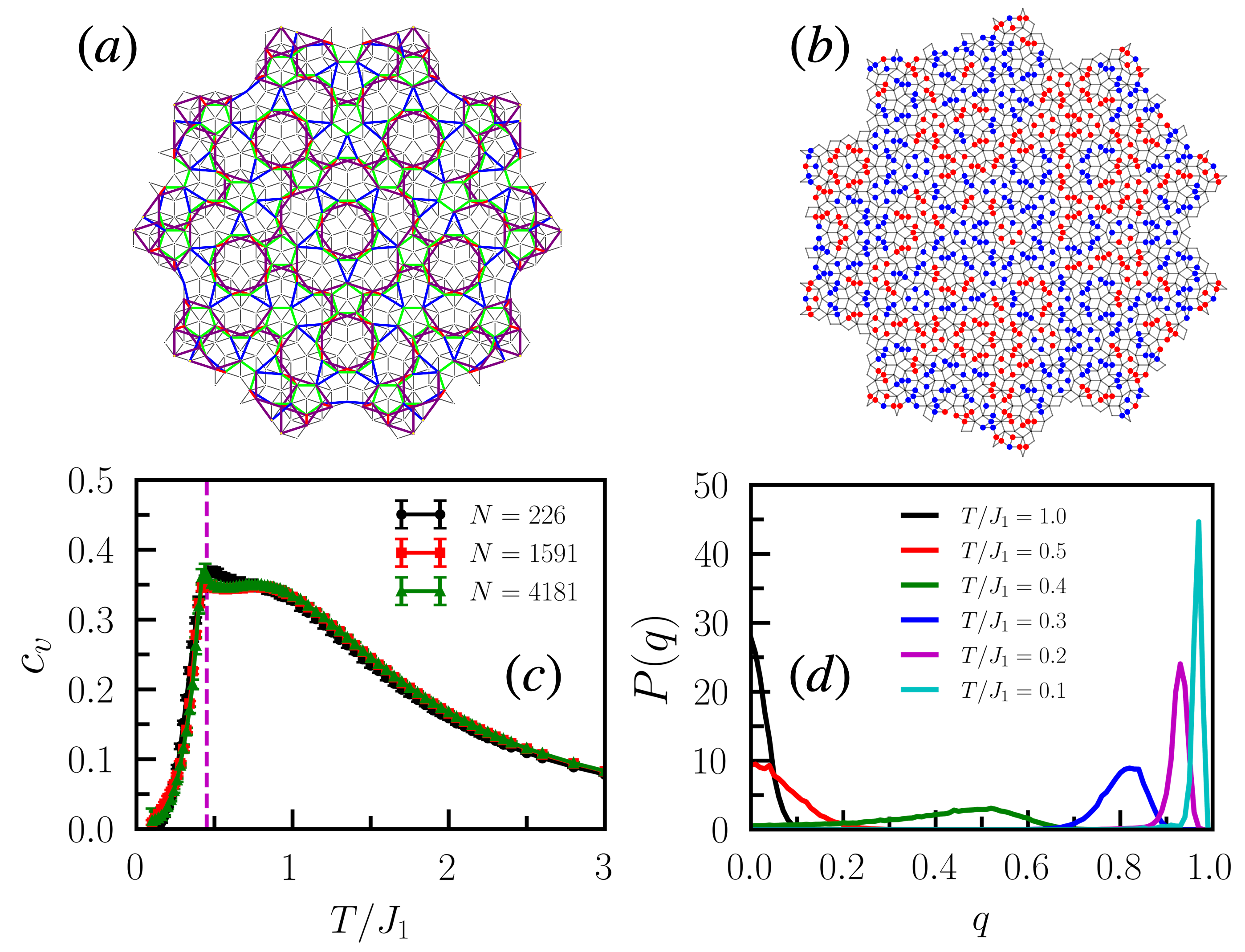}\caption{\label{fig:penrose_order}Monte Carlo results for local moments placed
at $z=3$ local environments in the Penrose lattice. (a) Minimal set
of exchange couplings for the $N=1591$ tiling with four magnetic
exchange couplings: $J_{1}$ (red), $J_{2}$ (blue), $J_{3}$ (green)
and $J_{4}$ (magenta). (b) Ground state configuration for $N=1591$,
$n=7/8$, and $J_{2}=-0.308J_{1}$, $J_{3}=0.442J_{1}$, $J_{4}=0.028J_{1}$.
The different colors indicate opposite spin orientations. (c) Specific
heat as a function of temperature for different approximant sizes.
The vertical dashed line marks the ordering temperature's position
$T_{N}=0.45\left(1\right)J_{1}$. (d) Histogram of the Edwards-Anderson
order parameter $P\left(q\right)$ for different temperatures $T$. }
\end{figure}

To gauge the generality of our results, we study Eq. \eqref{eq:ising}
in the Penrose tiling.  Again,  open boundary conditions are considered.
We show sample results in Fig. \ref{fig:penrose_order} where we consider
$n=7/8$ and the spins placed at sites with $z=3$ neighbors. The
minimal set of exchange couplings is displayed in Fig. \ref{fig:penrose_order}(a),
with the ground state illustrated in Fig. \ref{fig:penrose_order}(b).
We also obtain an intricate ordered ground state, and although the specific
heat in Fig. \ref{fig:penrose_order}(c) does not show a well-defined
peak, the residual entropy is tiny. We can confirm it first by studying
$P\left(q\right)$ in Fig. \ref{fig:penrose_order}(d).  For $T>T_N$,  $P\left(q\right)$ is peaked around $q=0$. 
As the temperature decreases,  the peak of $P\left(q\right)$ moves towards $q\approx1$. 

\begin{figure}[t]
\centering{}\includegraphics[width=1\columnwidth]{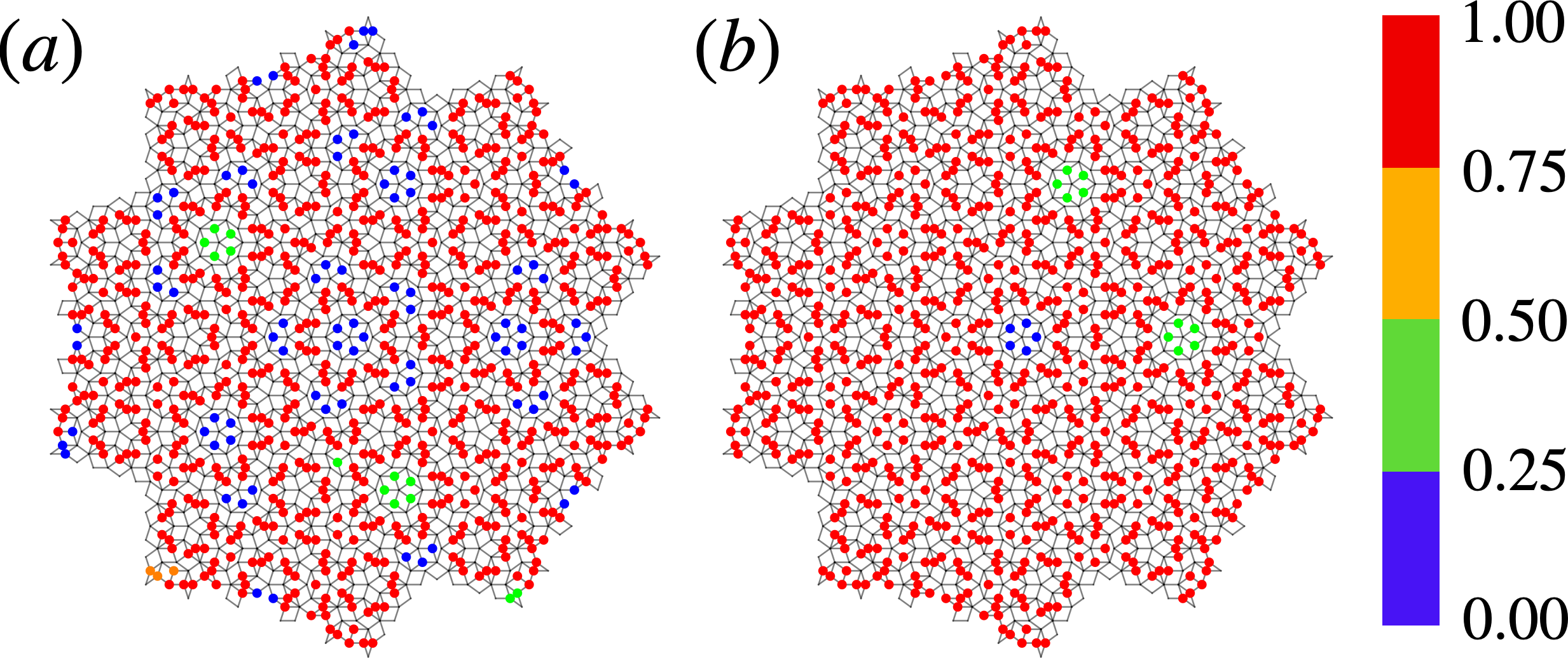}\caption{\label{fig:penrose_fluc}Snapshots of the spatial distribution of
the Edwards-Anderson order parameter $q_{i}$
for local moments placed at sites with $z=3$ local environments in
the Penrose tiling: (a) $T=0.1J_{1}<T_{N}$ and (b) $T=0.01J_{1}<T_{N}$.
We consider the same parameters as in Fig. \ref{fig:penrose_order}.}
\end{figure}

We show snapshots of the system configurations in Fig. \ref{fig:penrose_fluc}
at temperatures below $T_{N}$.  Spins surrounding
sites with $z=5$ fluctuate inside the ordered phase, but their number
is strongly suppressed as $T \to 0$,  consistent with $s_0 \approx 0$.  Although their contribution
to the thermodynamic response of the current system is minor,  it is worthwhile to note 
that the existence of partial order is a
robust and general phenomenon for local moments diluted in metallic
quasicrystals.  On the other hand,  it points to the general conclusion that the combination of deterministic quasiperiodicity and frustration is not enough to generate a spin-glass state within our minimal model and that true randomness is required. 

\section{\label{sec:epr}Local moment relaxation rate}

Ref. \citep{cabrera-baez19} presents a detailed ESR study in of rare-earth-based
quasicrystals and their approximants. ESR is a powerful technique
to probe the local electronic environment of diluted local moments.
It accesses the low-frequency transverse magnetization fluctuations
of the conduction electrons in the presence of a static magnetic field
\citep{barnes81}.

The reported ESR spectra in Ref. \citep{cabrera-baez19} displays
Dysonian line shapes, characteristic of localized moments embedded in
a conduction electron sea, thus giving access to the non-trivial electronic
properties of quasiperiodic systems. In particular, the linewidth
is linked to the local moment relaxation rate, $1/T_{1}^{i}$, which
is given by 
\begin{equation}
\frac{1}{T_{1}^{i}T}\propto\lim_{\omega_{0}\to0}\frac{\chi_{ii}^{\prime\prime}\left(\omega_{0}\right)}{\omega_{0}}\propto\int d\omega\,\rho_{i}^{2}\left(\omega\right)\left(-\frac{\partial f}{\partial\omega}\right),\label{eq:1t1}
\end{equation}
where $\chi_{ii}^{\prime\prime}\left(\omega_{0}\right)$ is the imaginary
part of the dynamical susceptibility of the conduction electrons,
$\omega_{0}$ is the ESR resonance frequency, $\rho_{i}\left(\omega\right)$
is the local density of states, and $f$ is the Fermi-Dirac distribution.
In a regular metal, $\rho_{i}\left(\omega\right)$ is roughly constant
around the Fermi level and we get a constant $1/T_{1}T$ at low temperatures,
a behavior known as Korringa relaxation law \citep{coleman15}. Given
the complex nature of the electronic states in a quasicrystal, it
is an interesting question to check if this law holds both in the
approximant and in the thermodynamic limit of the quasicrystal because
it is not obvious that one can neglect the generically strong fluctuations
of $\rho_{i}\left(\omega\right)$ as a function of $\omega$.

Modeling the conduction electrons with Eq. \eqref{eq:tbh}, we calculate
the local relaxation in Eq. \eqref{eq:1t1} rate using KPM since it
gives us direct access to the local density of states. For diluted
impurities, we do not select a local environment with a given $z$
and compute $\left\langle 1/T_{1}\right\rangle =N^{-1}\sum_{i=1}^{N}1/T_{1}^{i}$
considering all sites in the approximant. We did check that the general
conclusions we present hold for averages taken over a subset of sites
with a given $z$,  as expected by the inflation symmetry.

\begin{figure}[t]
\centering{}\includegraphics[width=1\columnwidth]{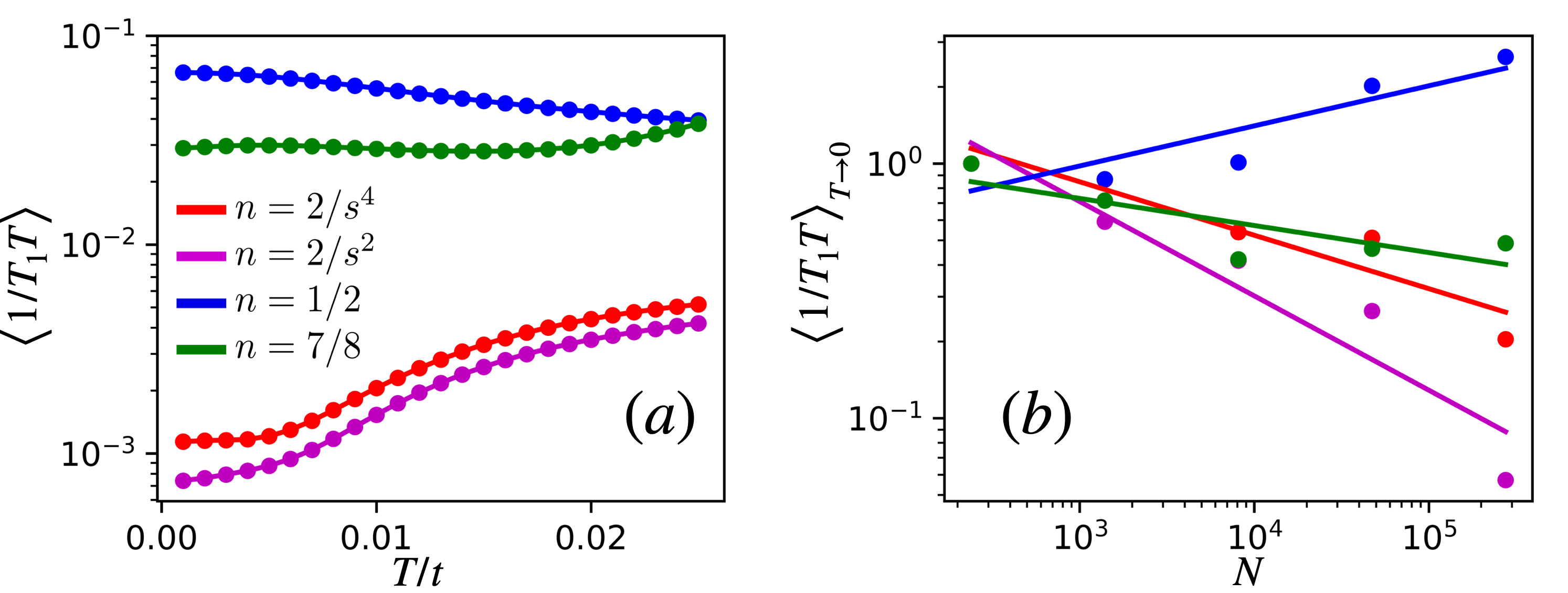}\caption{\label{fig:esr}Average spin relaxation rate, $\left\langle 1/T_{1}\right\rangle $,
divided by $T$, in a metallic quasicrystal. (a) $\left\langle 1/T_{1}T\right\rangle $
as a function of $T$ for the approximant $N=275807$ and several fillings $n$ in a mono-log scale. (b) Scaling of $\left\langle 1/T_{1}T\right\rangle $,
in the limit $T\to0$, as a function of the approximant size in a
log-log scale. We normalize the relaxation rates for different $N$
by their values for $N=239$. We use the same color scheme as in (a)
and fit the data by a power-law: $\left\langle 1/T_{1}T\right\rangle _{T\to0}\sim N^{-\alpha}$,
$\alpha=0.21\left(4\right)$, $0.37\left(7\right)$, $-0.16\left(5\right)$,
and $0.11\left(4\right)$ for $n=2/s^{4}$, $2/s^{2}$, $1/2$, and
$7/8$, respectively.}
\end{figure}

In Fig. \ref{fig:esr}(a) we show $\left\langle 1/T_{1}T\right\rangle $
as a function of the temperature for the $N=275807$ approximant.
For all fillings considered, we recover the Korringa relaxation law
at low temperatures, $T\lesssim0.005t$. The fillings $n=2/s^{4}$
and $n=2/s^{2}$ correspond to pseudogaps in the density of states
for $N\to\infty$. This amounts for the small values of $\left\langle 1/T_{1}T\right\rangle $
for those two fillings in comparison to the more generic fillings
$1/2$ and $7/8$. Actually, for $n=2/s^{2}$, $\left\langle 1/T_{1}T\right\rangle $
still shows a mild temperature dependence down to $T\to0$.

Extrapolating the curves $\left\langle 1/T_{1}T\right\rangle $ down
to $T=0$, we plot $\left\langle 1/T_{1}T\right\rangle _{T\to0}\times N$.
We normalize $\left\langle 1/T_{1}T\right\rangle _{T\to0}$ by its
value at $N=239$ to account for the differences in scale coming from
$\rho\left(\omega\right)$. The curves are well fitted by a power
law, $\left\langle 1/T_{1}T\right\rangle _{T\to0}\sim N^{-\alpha}$.
For the fillings $n=2/s^{4}$ and $n=2/s^{2}$, the decay is sharper
because of the expected pseudogap in the thermodynamic limit. For
$n=7/8$, there is a milder decay, indicating that $\rho\left(\omega\right)$
is reduced with $N$ for this particular filling. Interestingly, for
$n=1/2$ $\rho\left(\omega\right)$ increases and $\alpha<0$. Since
the integral of $\rho\left(\omega\right)$ over all states is $2$,
the decrease at a given filling must be compensated by an increase
elsewhere. Most of this enhancement comes from $n=1$ at $\omega=0$
where the existence of localized states imply that $\rho\left(\omega\to0\right)\to\infty$
for $N\to\infty$ \citep{rieth95}.

Overall, a generic filing might increase or decrease the value of
the relaxation rate as one approaches the thermodynamic limit. Such
a finite-size study is not feasible in a periodic solid-state system.
However, one can contrast the behavior of approximant and the thermodynamic
quasicrystal in a quasiperiodic system,  as one controls the size of the unit cell \citep{cabrera-baez19}. If
these results are interpreted according to the general scenario suggested
by Fig. \ref{fig:esr}, one might gain valuable information regarding
the non-trivial electronic structure of quasicrystals.

\section{\label{sec:conclusions}Conclusions}

Motivated by experimental studies of magnetic order in rare-earth
metallic quasicrystals \citep{goldman13,cabrera-baez19,shiino21},
we followed Refs. \citep{thiem15a,thiem15b} and studied local Ising
moments placed in the two-dimensional Ammann-Beenker and Penrose tiling. Modeling
the electronic properties of this quasiperiodic system via a tight-binding
model, we investigated the resulting ordering states for local moments
interacting via a minimal set of RKKY-like interactions \citep{duneau91}.
Taking advantage of the fact that in a quasicrystal we have different
local environments, we place the local moments in a deterministic
fashion: the spins occupy all sites with a given local coordination
number $z$.

Monte Carlo simulations in finite tilings, combined with usual finite
size scaling, show that the system generically develops long-range
order at low temperatures \citep{thiem15a,thiem15b}. Therefore, the
lack of periodicity and local moments interacting via RKKY couplings
are insufficient to generate a spin-glass, and true randomness is
required. This scenario is broadly consistent with the experimental
results, for instance, in Refs. \citep{goldman13,cabrera-baez19,shiino21},
reporting long-range order in rare-earth-based metallic quasicrystalline
approximants. Our results suggest these magnetic states to be quite
complex and fragile, owing to the presence of quasiperiodicity and
frustration. In particular, a fraction of the sites might fluctuate to low
temperatures since they feel no mean field due to the ordered spins.
One experimental signature of such a partially disordered state is
a residual ground state entropy, which should be quenched in external
fields.

Due to these disordered spin clusters, perturbations substantially
impact the ordered state. In particular, we show that the introduction
of vacancies turns this ordered state of the approximant into a spin-glass
\citep{shiino21}.

For the quasicrystal, one might envision the following situation.
Consider a cluster of fluctuating spins of size $D$, which is weakly
coupled to the ordered local moments. According to Conway's theorem
\citep{gardner77}, such a local pattern is never more than a distance
$2D$ of a precisely identical region. This implies a finite fraction
of spins experiencing weak mean fields from the ordered local moments
in the limit $N\to\infty$. An extrinsic perturbation can thus easily
pin these spins into random directions, effectively generating a random
field on the ordered spins that will melt the long-range order \citep{aharony78,michel21, ye22}.
Such a mechanism could help explain why experiments in magnetic quasicrystals
generically identify spin-glass order.

The ESR experiments of Ref. \citep{cabrera-baez19} report that, in
the diluted limit, the local moment relaxation rate $1/T_{1}T$ obeys
the Korringa relaxation law, both for the approximant and the quasicrystal.
However, $1/T_{1}T$ diminishes as one moves from the approximant
to the quasicrystal. Within our simple model, one could interpret
this result assuming that the Fermi level of the conduction electrons
sits close to a pseudogap, as it is often the case with metallic quasicrystals
\citep{fujiwara89,fujiwara91,jazbec14}.  However, further investigation is required, for instance, by performing transport \citep{pierce93,fisher99} and spectroscopy measurements \citep{rogalev15}.

In conclusion, we studied the behavior of local moments in contact
with a quasiperiodic conduction electron bath. The spin-relaxation
rate in the diluted limit depends on the system size $N$ and the
electronic filling $n$. If the Fermi energy sits at a pseudogap
for $N\to\infty$, the relaxation rate is suppressed with system size.
For a finite concentration of magnetic impurities, we investigated
the resulting ordering pattern if these spins interact via RKKY-like
interactions, mediated by the electronic bath. For impurities occupying
all sites with a fixed number of nearest neighbors in the approximants, we generically find
long-range magnetic order for finite $N$. However, the resulting
magnetic state is generically fragile, usually displaying clusters
of essentially free spins. As $N\to\infty$, these clusters become
increasingly sensitive to extrinsic defects and could nucleate random
fields destabilizing the long-range magnetic order, paving the way
for a spin-glass state. Local probes investigating the local distribution
of fields, like NMR, could be useful in testing this scenario. 
\begin{acknowledgments}
We acknowledge useful discussions with M. Ávila, V. Dobrosavljevi\'{c},
M. Cabrera-Baez, A. Jagannathan, and M. Vojta.  We thank the anonymous referees for the insightful referring process that markedly improved our manuscript.  R.N.A was supported
by the CAPES (Brazil) - Finance Code 001. C.C.B was supported by FAPESP
(Brazil), Grants No. 2022/14330-1 and 2023/06101-5. ECA was supported
by CNPq (Brazil), Grant No. 302823/2022-0, and FAPESP (Brazil), Grants
No. 2021/06629-4 and 2022/15453-0. 
\end{acknowledgments}


\appendix

\section{\label{sec:further}Further exchange couplings}

We now present results including extra exchange couplings in Eq. \eqref{eq:ising}.
Specifically, we consider $J_{5\left(4\right)}$ between fifth (fourth)
neighbor for impurities placed in the $z=3\left(4\right)$ local environment of the octagonal tiling.

\begin{figure}[t]
\centering{}\includegraphics[width=1\columnwidth]{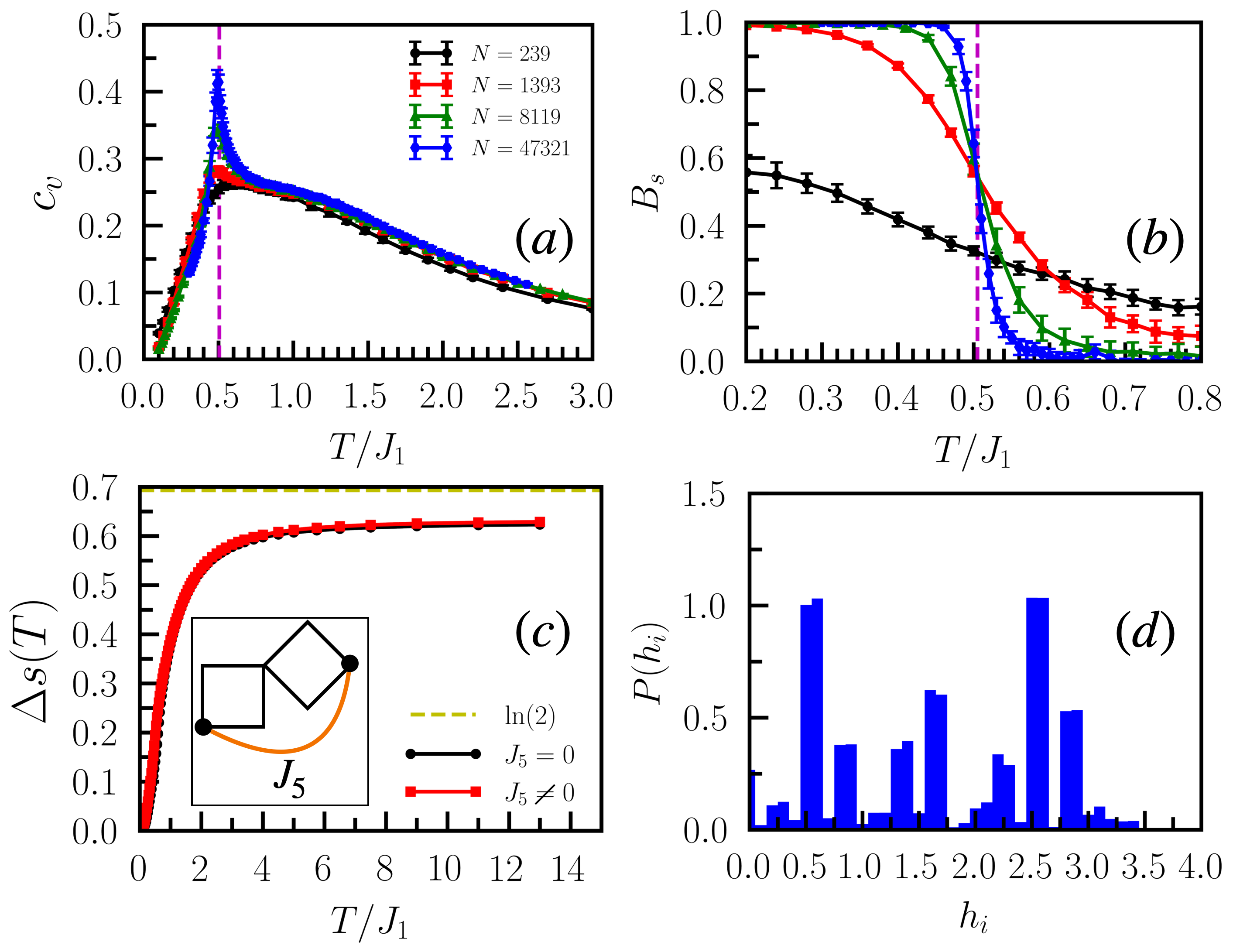}\caption{\label{fig:z3_j5}Characterization of the ordered states for local
moments placed at the $z=3$ sites of the Ammann-Beenker tiling for $n=2/s^{2}$ and an extra coupling
$J_{5}=-0.142J_{1}$, see the inset of (c). (a) Specific heat as a
function of the temperature $T$ for several approximant sizes $N$.
The vertical dashed line marks the position of the ordering temperature
$T_{N}$. (b) Binder cumulant $B_{s}$ of the order parameter as
a function of $T$. The dashed line marks the crossing of the curves
for $N\ge1393$ and determines $T_{N}=0.505\left(5\right)J_{1}$.
The curve colors are the same as in (a). (c) Residual entropy $\Delta s\left(T\right)$,
per site, as a function of $T$ for $N=8119$. The dashed line gives
$s\left(T\to\infty\right)=\log{2}$. (d) Histogram of the local exchange
field, $P\left(h_{i}\right)$, for $T<T_{N}$. }
\end{figure}

For $z=3$, we consider $n=2/s^{2}$ and the sample results are displayed
in Fig. \ref{fig:z3_j5}. Despite a ferromagnetic $J_{5}$, see Fig.
\ref{fig:z3_j5}(c), the critical temperature is reduced to $T_{N}=0.505(5)J_{1}$,
indicating an enhancement in frustration. However, the results are qualitatively
similar, as shown by the residual entropy $s_{0}$ and the local exchange
field distribution for $T<T_{N}$, Figs. \ref{fig:orderz3}(c) and
~\ref{fig:orderz3}(d).  Because $P\left(h_{i}\to0\right)\neq0$ and $s_{0}$ hardly change,
we conclude that longer-range couplings do not quench the fluctuating
spin islands for $T\to0$. This conclusion agrees with Refs. \citep{thiem15a,thiem15b}.  
These references report fluctuating spins even when taking into account exchange
couplings between all spin pairs as determined by the full set of RKKY interactions. 

\begin{figure}[t]
\centering{}\includegraphics[width=1\columnwidth]{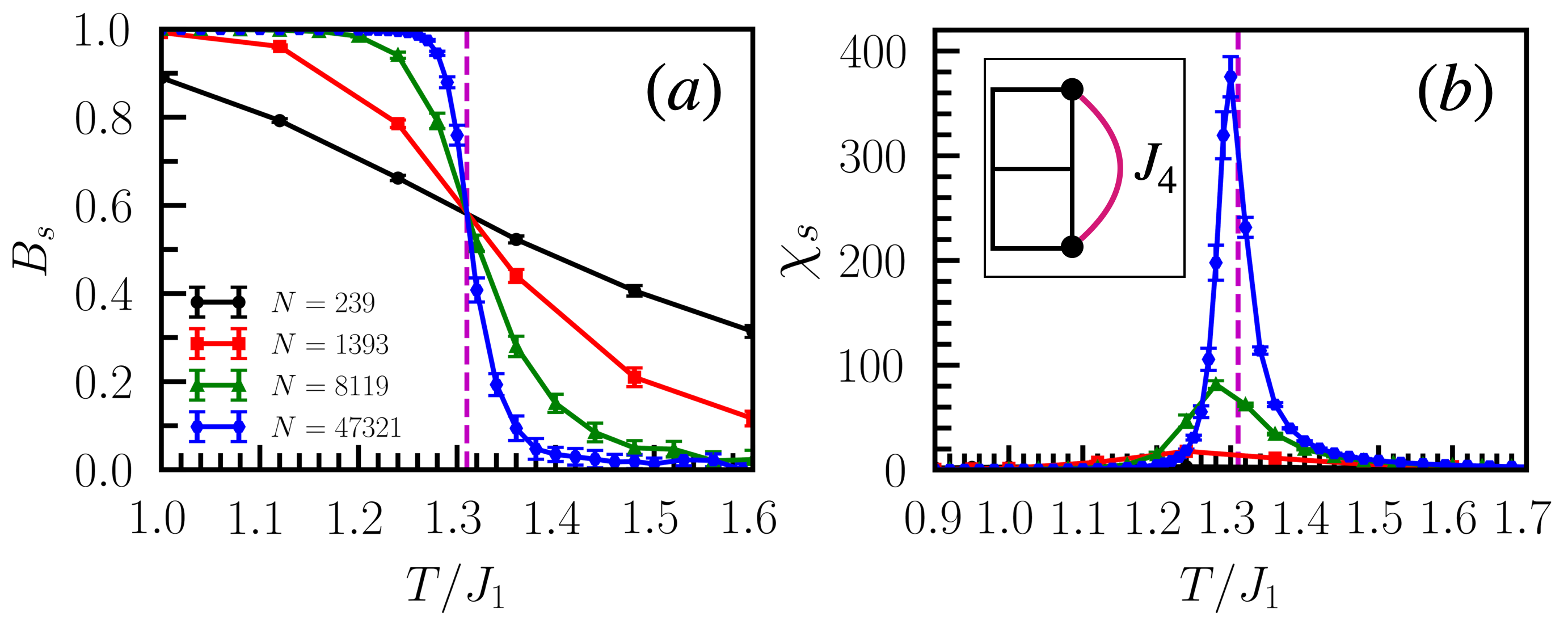}\caption{\label{fig:z4_j4}Characterization of the ordered states for local
moments placed at the $z=4$ sites, $n=7/8$ and an extra coupling
$J_{4}=-0.077\left|J_{1}\right|$, see the inset of (b). (a) Binder
cumulant $B_{s}$ of the order parameter as a function of $T$.
The dashed line marks the crossing of the curves and determines $T_{N}=1.31\left(1\right)\left|J_{1}\right|$.
(b) Order parameter susceptibility $\chi_{s}$ as a function of $T$.
The vertical dashed line marks the position of the ordering temperature
$T_{N}$. The curve colors are the same as in (a).}
\end{figure}

For $z=4$, we consider the filling $n=7/8$ and a fourth
neighbor exchange $J_{4}$, see Fig. \ref{fig:z4_j4}(d), which is
again ferromagnetic. The critical temperature slightly increases to
$T_{N}=1.31\left(1\right)\left|J_{1}\right|$, but the results hardly
change with respect to Fig. \ref{fig:z4_MC_thermo}. 

As longer-range couplings tend to diminish in magnitude, see Fig.
\ref{fig:RKKY}, we conclude that further couples do not alter qualitatively
the scenario presented in the main text. Therefore, keeping the minimal
set of $J_{i}^{\prime}$s is sufficient to capture the key aspects
of the model. 

%

\end{document}